\documentclass[a4paper,12pt]{article}
\usepackage[utf8]{in
    putenc}
\usepackage{cancel}
\usepackage{ulem}
\usepackage{amsfonts}
\usepackage{amssymb}
\usepackage{multicol}
\usepackage{graphicx}
\usepackage{amsmath,bm}
\usepackage{enumerate}
\usepackage{mathtools}
\usepackage{tikz} \usetikzlibrary{calc}
\usepackage[countmax]{subfloat}
\usetikzlibrary{fit}
\usepackage{tabularray}
\usepackage{longtable}
\usepackage{xcolor}
\usepackage{float}
\usepackage{color}
\usepackage{here}
\usepackage{cite}
\usepackage{subcaption}
\usepackage{mwe}
\usepackage[outline]{contour}
\usepackage{tabularx}
\usepackage{multirow}
\usepackage{mathrsfs}
\usepackage{float,epsfig}
\usepackage{dcolumn}% Align table columns on decimal point
\usepackage{pgfplots}
\usepackage{array}
\pgfplotsset{compat=1.18}
\usepackage{graphicx}% Include figure files
\usepackage{bm}% bold math
\usepackage{amsmath,amssymb,amsthm}
\usepackage[colorlinks=true,linkcolor=blue,citecolor=red]{hyperref}
\definecolor{lime}{HTML}{A6CE39}
\newcommand{\orcidicon}{%
    \begin{tikzpicture}
    \draw[lime, fill=lime] (0,0)
        circle [radius=0.16]
        node[white] {{\fontfamily{qag}\selectfont \tiny ID}};
    \draw[white, fill=white] (-0.0625,0.095)
        circle [radius=0.007];
    \end{tikzpicture}   \hspace{-2mm}
}
\newcommand\orcidHasan{{\href{https://orcid.org/0000-0001-7408-0910}{\orcidicon}}}
\newcommand\orcidKarima{{\href{https://orcid.org/0000-0001-5419-8516}{\orcidicon}}}
\newcommand\orcidFaical{{\href{https://orcid.org/0000-0002-2977-0821}{\orcidicon}}}
\title{\bf Thermal chaos of charged-flat black hole via Rényi formalism}

\author{
F. Barzi\orcidFaical\!\!$^{1,3}$\thanks{faical.barzi@edu.uiz.ac.ma},  
 H.  El Moumni\orcidHasan\!\!$^1$\thanks{h.elmoumni@uiz.ac.ma (Corresponding author)}, K. Masmar\orcidKarima\!\!$^{1,2}$\thanks{karima.masmar@gmail.com}
\\
{\small $^{1}$ LPTHE, Physics Department, Faculty of Sciences, Ibnou Zohr University, Agadir, Morocco. }\\
{\small $^{2}$Laboratory of  High Energy Physics and Condensed Matter
HASSAN II University,}\\{\small Faculty of Sciences Ain Chock, Casablanca, Morocco.}\\
{\small $^{3}$CRMEF, Regional Center for Education and Training Professions Marrakesh, Morocco.
}
}

\date{\today}

\begin{document} 
	\maketitle
\begin{abstract}

%Charged-flat black holes in Rényi extended phase space exhibit phase structures that parallel those of a van der Waals fluid in four-dimensional spacetime, and the Reissner-Nordstrom-Anti-de-Sitter black holes in the standard Gibbs-Bomtzmann extended phase space. In this paper, we explore the dynamics of states initially within the unstable spinodal region of phase space, subjected to time-periodic thermal perturbations through the Mel'nikov method. Our findings indicate that chaos emerges when $\delta$ exceeds the critical value of $\delta_c$. Additionally, we investigate space-periodic thermal perturbations on its equilibrium state, finding that chaos is invariably present, regardless of perturbation amplitude.
Charged-flat black holes in the Rényi extended phase space demonstrate phase structures akin to those of a van der Waals fluid in four-dimensional spacetime and mirror the behaviors of Reissner-Nordstrom-Anti-de-Sitter black holes within the standard Gibbs-Boltzmann extended phase space. This study delves into the dynamics of states initially positioned within the unstable spinodal region of the phase space associated with the charged-flat black hole when subjected to time-periodic thermal perturbations. Our analysis based on the Mel'nikov method reveals that chaos emerges when the $\delta$ parameter surpasses a critical threshold, $\delta_c$. This critical quantity is dependent on the black hole charge; notably, a larger value of $Q$ impedes the onset of chaos.

Furthermore, we examine the effects of space-periodic thermal perturbations on its equilibrium state and find that chaos invariably occurs, irrespective of the perturbation amplitude. Hence, the chaotic dynamics observed in the analysis of charged-flat black holes under Rényi statistics exhibit resemblances to those of asymptotically AdS-charged black holes investigated via the Gibbs-Boltzmann formalism. 
This serves as yet another example of a potential and significant connection between the cosmological constant and the nonextensivity Rényi parameter.
		{\noindent }
\end{abstract}
	%\addcontentsline{toc}{section}{\nameref{appendix}}
	\tableofcontents
	
	%\newpage
	%\newpage

	%	 \tableofcontents

\section{Introduction}

Investigations into the phase transitions of black holes across diverse backgrounds, particularly within Anti-de Sitter (AdS) space-time, have captured considerable interest, notably within the realm of black hole thermodynamics. This exploration is driven by gravitational considerations and motivated by holographic principles \cite{Christodoulou:1970wf,Bekenstein:1973ur,Bekenstein:1974ax,Hawking:1975vcx,Hawking:1976de,Gibbons:1976ue,Hawking:1982dh,Maldacena:1997re,Witten:1998qj,Gubser:1998bc}. Recently, a fresh perspective has emerged, treating the cosmological constant as a dynamically variable entity analogous to pressure. This conceptual shift has led to the development of extended phase thermodynamics, altering the conventional first law of black hole mechanics through the addition of a new $pdV$ term \cite{Kastor:2009wy,Caldarelli:1999xj,Wang:2006eb,Sekiwa:2006qj,Dolan:2010ha,Cvetic:2010jb,Henneaux:1984ji,Teitelboim:1985dp,Henneaux:1989zc,Johnson:2018amj}. By scrutinizing the $P-V$ critical behavior across various black hole configurations, researchers have uncovered a compelling correspondence between the small/large phase transitions of black holes and the behavior exhibited in the Van der Waals liquid/gas system \cite{Chamblin:1999tk,Chamblin:1999hg,Kubiznak:2012wp,Kubiznak:2014zwa,Kubiznak:2016qmn,Belhaj:2015hha,Chabab:2019kfs,Belhaj:2020nqy}. This interdisciplinary fusion underscores the intricate interplay between gravitational physics, holography, and the fundamental principles governing high-energy phenomena, offering profound insights into the nature of spacetime and its intricate relationship with thermodynamic concepts at the forefront of physical understanding.

Recent studies \cite{Czinner:2015eyk,Biro:2011ncf}, have brought into question related to the applicability of standard Gibbs-Boltzmann (GB) statistics, particularly in systems with long-range interactions such as black holes. Moreover, authors \cite{Promsiri:2020jga,Barzi:2022ygr,Nakarachinda:2021jxd,Li:2020khm,Cimidiker:2023kle,Chunaksorn:2022whl,Cimdiker:2022ics,Hirunsirisawat:2022fsb,Sriling:2021lpr,Nojiri:2021czz,Promsiri:2021hhv,Samart:2020klx} have highlighted the limitations of traditional stability analyses in the context of black hole systems, attributing these challenges to the non-additivity of entropy, as observed in the Bekenstein-Hawking entropy. These issues collectively indicate that employing GB statistics within self-gravitating systems may yield incomplete results.

 A fundamental aspect complicating the application of GB statistics is the nonextensivity of black hole entropy. Unlike conventional systems where entropy scales with volume, black hole entropy is directly proportional to its surface area. This discrepancy underscores the necessity of exploring black hole thermodynamics through the lens of nonextensive statistics and entropy.

Fortunately, an alternative approach emerges with the introduction of a weaker composition rule \cite{PhysRevE.63.061105}, notably Abe's one
\begin{equation}\label{Abe_rule}
H_\lambda(S_{AB})=H_\lambda(S_A)+H_\lambda(S_B)+\lambda H_\lambda(S_A)H_\lambda(S_B),
\end{equation}
in which $H$ denotes a differentiable function of entropy $S$,  $\lambda$ is a real parameter, while, $A$ and $B$ are two independent systems. One of nonextensive entropies, Tsallis entropy \cite{Tsallis:1987eu,Tsallis:2012js}
takes the form
\begin{equation}\label {abe}
S_q=\frac{1-\sum_{i=1}^\Omega p_i^q}{q-1}\equiv S_{T},\quad q\in \mathbb{R}.
\end{equation}
Here, $\Omega\in\mathbb{N}$, is the total number of possible configurations and $p_i$ is the associated probabilities. The standard GB entropy $S_{GB}=-\sum_{i=1}^\Omega p_i\ln p_i$ can be easily recovered by setting $q\to 1$. In this  framework, by setting $H_\lambda(S)=S_T$ and $\lambda=1-q$ the pseudo-additive rule becomes
\begin{equation}\label{biro_van}
S_T(A\cup B)=S_T(A)+S_T(B)+(1-q)S_T(A)S_T(B).
\end{equation}
Here $A$ and $B$ are two thermodynamic subsystems. Within such pseudo-additivity form, we fail to define an empirical temperature. To resolve this issue Biró and Van \cite{Biro:2011ncf} suggest the introduction of the formal logarithm of the entropy and which corresponds to the Rényi one 
\begin{equation}\label{Renyi_def}
L(S_T)=\frac{1}{1-q}
\ln\left[1+(1-q)S_T\right]\equiv S_R. 
\end{equation}

This parameter was originally introduced to account for the non-additivity of Bekenstein-Hawking entropy expressed by \eqref{biro_van}, but in turn, endows the Rényi entropy with a remarkable additive composition law such as,
\begin{equation}
    S_{R}(A\cup B)=S_{R}(A)+S_{R}(B),\label{composition_Renyi}
\end{equation}
 which can be easily verified using Eqs.\eqref{biro_van} and \eqref{Renyi_def}. Therefore,\textit{ Rényi entropy resolves the non-extensive nature of black holes while retaining the additivity of their entropy}.

Moreover, the Rényi entropy $S_R=\frac{1}{\lambda}\ln(1+\lambda S_q)$ is compatible  with the zero law of thermodynamics and allows a well-defined form the empirical temperature
\begin{equation}
\frac{1}{T_R}=\frac{\partial S_R (E)}{\partial E},
\end{equation}
where $E$ stands for the energy of the system. This entropic alternative framework offers promise in addressing the challenges posed by nonextensive systems, providing a pathway to more accurately model black hole thermodynamics within the constraints of nonextensivity.

The inevitability of chaos in certain natural dynamical systems, including those found in black hole physics and cosmology, has been well-documented \cite{Bombelli:1991eg,Letelier:1996he,Santoprete:2001wz,Monerat:1998tw,Felder:2006cc,Gair:2007kr,Chen:2016tmr}. Past investigations have delved into the chaotic behavior of black holes utilizing various methodologies, ranging from the computation of Lyapunov exponents to the study of stability in orbits and quasinormal modes in different black hole geometries \cite{Cardoso:2008bp,Konoplya:2017wot}. Mel'nikov's method\cite{Melnk:1963jan}, originally developed within the context of dynamical systems, has also been employed to explore chaotic behavior in the motion of geodesics \cite{Bombelli:1991eg,Letelier:1996he}. However, the exploration of chaos within the realm of black hole thermodynamics and phase transitions has only recently begun to emerge \cite{Chabab:2018lzf}. This nascent field owes its growth in part to recent developments linking black hole thermodynamics with the behavior of Van der Waals systems, facilitated by the inclusion of a pressure term in the first law \cite{Chamblin:1999hg,Kubiznak:2012wp,Kubiznak:2014zwa}.

In our previous study \cite{Chabab:2018lzf}, the Mel'nikov method, extensively used in dynamical systems, was applied to investigate chaos in the context of black holes within extended phase space. This novel approach, inspired by previous applications in the study of Van der Waals systems, introduced temporal and spatial period perturbations into the $P-v$ thermodynamic phase space. The presence of chaos was discerned through the analysis of zeros of the Mel'nikov function. Moreover, a notable discovery was the establishment of a bound tied to the charge of the black hole, beyond which the system exhibits chaotic behavior. This groundbreaking research paves the way for deeper explorations into the intricate interplay between chaos and black hole thermodynamics \cite{Chen:2019bwt,Mahish:2019tgv,Dai:2020wny,Tang:2020zhq}, shedding light on fundamental aspects of the universe's dynamics.

This study aims to make a valuable contribution to this area of research, driven by the non-extensive characteristics of black hole entropy and the presence of a Van der Waals-like phase structure in the context of chaotic dynamics within charged flat black hole backgrounds.

\section{The charged-flat black hole Rényi thermodynamics: a concise review }

	The first step suggests revisiting the four-dimensional Einstein-Maxwell action, $\mathcal{I}_\text{EM}$, which provides the framework for classical charged solutions in Einstein's general relativity. This formulation incorporates the metric tensor $g_{\mu\nu}$, possessing a metric signature of $(-,+,+,+)$, and the $1$-form gauge field $A = A_{\mu}dx^{\mu}$, where $\mu, \nu = 0, 1, 2, 3$. The action is obtained to be
 \begin{equation}
     \mathcal{I}_\text{EM}=\frac{1}{16\pi G}\int_{}\: dx^4 \sqrt{-g}(\mathcal{R}-F^2).
 \end{equation}
 Where, $\mathcal{R}=\mathcal{R}^{\mu\nu}\mathcal{R}_{\mu\nu}$ is the \textit{Ricci} scalar, $g=det(g_{\mu\nu})$ denotes the determinant of the metric and the Maxwellian field strength $F$, is a closed 2-form defined as $F=dA=\frac{1}{2}F_{\mu\nu} dx^{\mu}\wedge dx^{\nu}$ with the tensor components $F_{\mu\nu}=\nabla_{\mu}A_{\nu}-\nabla_{\nu} A_{\mu}$. Minimizing the action, %\textit{The stationary action principle}, 
 $\delta \mathcal{I}_{EM}=0$, permits to derive the Euler-Lagrange equations of motion as %known as \textit{Einstein-Maxwell field equations}
 \begin{eqnarray}\nonumber
    &\nabla^\mu  F_{\mu\nu}=0,\label{eq_EM2}\\ &\nabla_\rho  F_{\mu\nu}+\nabla_\mu  F_{\nu\rho}+\nabla_\nu  F_{\rho\mu}=0,\label{eq_EM3}\\
    &R_{\mu\nu}=2F_{\mu \alpha}F_{\nu}^{\alpha}-\frac{1}{2}F^2 g_{\mu\nu}.\label{eq_EM1}
     \nonumber
 \end{eqnarray}
We note that a gravitational normalization scheme is adopted for the field $F$ in the previous equations %Eqs.\eqref{eq_EM2} 
to absorb a factor of $8\pi G$. 
For a spherically symmetric, \textit{asymptotically flat Reissner-Nordström black hole}, characterized by mass $M$ and electric charge $Q$ where $M \geq |Q|$, often termed the \textit{charged-flat black hole}, the only non-trivial components of the anti-symmetric field strength $F$ are
\begin{equation}
    F_{tr}=-F_{rt}.
\end{equation}
%Furthermore, the most general static and spherically symmetric ansatz in four dimensions has the following form
%\begin{equation}
 %   ds^2=-U(r)^2dt^2+V(r)^2dr^2+r^2d\omega^2_2.\label{metric_ansatz}
%\end{equation}
%In which, $d\omega^2_2 = d\Theta^2 + \sin^2\Theta d\Phi^2$ denotes the square of a line element on the 2-sphere, while $U$ and $V$ are coordinate functions. 
By solving the Einstein-Maxwell equations, %, Eqs.\eqref{eq_EM2} for the above metric, Eq.\eqref{metric_ansatz}, we find
%\begin{equation}
 %   U(r)=V(r)^{-1}=1-\frac{2M}{r}+\frac{Q^2}{r^2}\equiv f(r).
%\end{equation}
 the line element of the charged-flat black turns out to be
	%We first write the spherically symmetric Reissner-Nordström of mass $m$ and charge $Q$, the line element is given by
	\begin{eqnarray}
	ds^2 = -\left(1-\frac{2M}{r} + \frac{Q^2}{r^2}\right)dt^2 + \frac{dr^2}{\left(1-\frac{2M}{r} + \frac{Q^2}{r^2}\right)} + r^2(d\Theta^2 + \sin^2\Theta d\Phi^2),
	\end{eqnarray}
	where it has been recognized that the outer and inner horizons correspond to the roots of the blackening function, which are expressed as follows  
	\begin{eqnarray}
	r_{\pm}=M \pm \sqrt{M^2-Q^2}.
	\end{eqnarray}
	The black hole event horizon corresponds to the largest root $r_h=r_+$. This enables the expression of the mass $M$ in terms of the horizon radius $r_h$ and the charge $Q$ as
	\begin{eqnarray}
	M = \frac{r_h}{2}\left( 1 + \frac{Q^2}{r_h^2} \right). \label{mass}
	\end{eqnarray}
	While the charge $Q$  generates the bulk gauge field
	\begin{eqnarray}
	A = A_tdt = -\left(\frac{Q}{r}-\Phi \right)dt.  \label {bh4}
	\end{eqnarray}
	We can determine the electric potential $\Phi$ by setting $A_t=0$ at the horizon $r=r_h$. Thus, we find
	\begin{equation}
	\Phi = \frac{Q}{r_{h}} = \frac{Q}{M + \sqrt{M^2 - Q^2}}. \label {bh6}
	\end{equation}

\begin{figure}[!ht]
		\centering
		\begin{tabbing}
			\centering
			\hspace{-22mm}%\=\kill
			\includegraphics[scale=.42]{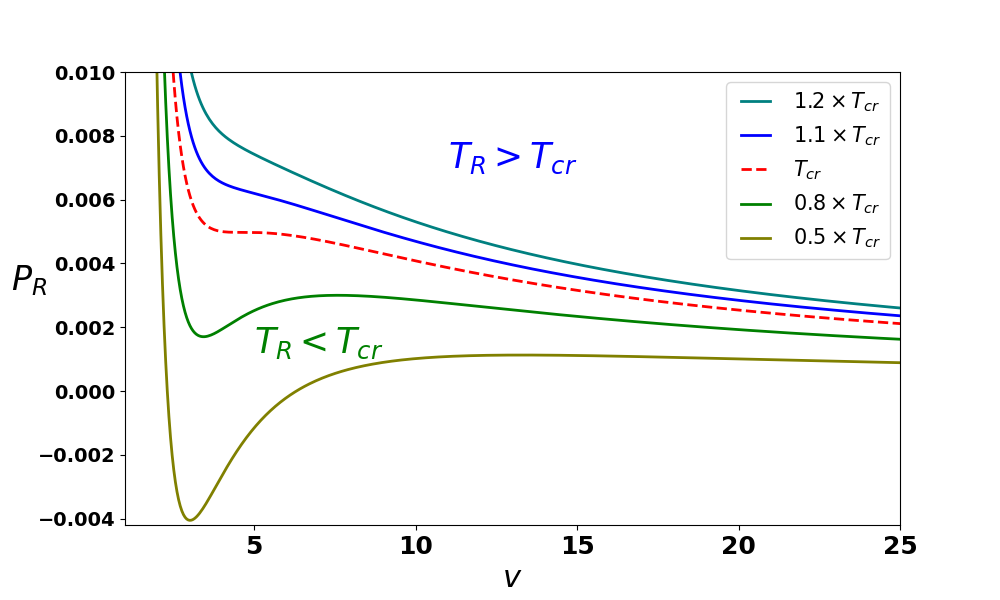}\label{PRenyi_V}%\>
			\hspace{-0.7cm}
			\includegraphics[scale=.42]{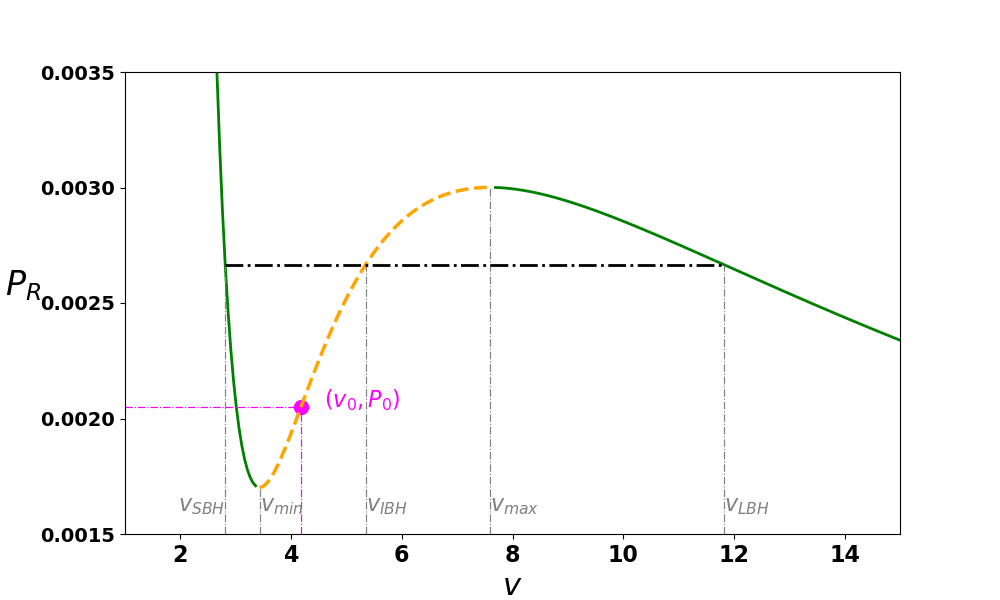}%\\
		\end{tabbing}
		\vspace{-0.7cm}
				\caption{\footnotesize\it \textbf{Left:} The $ P_R - v $ isothermal curves with different temperatures $ T_R $ for the charged-flat black holes with Q=$1/\sqrt{2}$: $ T \leq T_{cr} $ (green curves), $ T = T_{cr} $ (red dashed line) and $ T > T_{cr} $ (blue curves). \textbf{ Right:}, the subcritical curve can be divided into two stable regions (green lines) and an unstable spinodal one (orange dashed line). The black dashed line is the coexisting line of the large black hole (with specific volume $ v_{LBH} $) and a small black hole (with specific volume $ v_{SBH}$) with phase transition pressure $ P_0 $.}
		\label{fig:prenyiv}
	\end{figure}

As indicated in \cite{Promsiri:2020jga,dilaton}, the Rényi entropy $S_R$ is defined as the formal logarithm of the Bekenstein-Hawking entropy $S_{BH}$, treated as the Tsallis entropy $S_{T}$ in Eq. \eqref{Renyi_def}, thus
\begin{equation}
S_R=\frac{1}{\lambda}\ln(1+\lambda S_{BH}). \label{bh17}
\end{equation}
It is noteworthy that in the limit of a vanishing nonextensivity parameter $\lambda$, we regain the conventional Gibbs-Boltzmann statistics, denoted as $S_R\overset{\lambda\rightarrow0}{\longrightarrow}S_{BH}$. In this study, we proceed with the assumption that the nonextensivity parameter $\lambda$ is small and positive, specifically $0 < \lambda \ll 1$. This assumption is based on the notion that non-extensive effects serve as first-order corrections to the traditional framework of extensive statistical mechanics.  Further, introducing a first-order correction in $\lambda$ to the charged-flat black hole temperature. The Rényi temperature $T_R$ can be formulated as \cite{Promsiri:2020jga}
\begin{eqnarray}
T_R = \frac{1}{\partial{S_R/\partial{M}}} &=& T_H(1+\lambda S_{BH})\label{Tr}\\ 
&=& \frac{(r_h^{2}-Q^2)(1+ \lambda \pi r_h^{2})}{4\pi r_h^{3}}. \label{bh25}
\end{eqnarray}
where $T_H= \frac{r_h^{2}-Q^2}{4\pi r_h^{3}}$ and $S_{BH}=\pi r_h^{2}$ are the Hawking temperature and the Bekenstein-Hawking entropy of charged-flat black hole, respectively. 
Furthermore, within the Rényi extended phase space \cite{Promsiri:2020jga}, the nonextensivity parameter $\lambda$ is linked with the Rényi thermodynamic pressure, the electric potential $\Phi$, and the charge $Q$ through
\begin{eqnarray}\label{p-lam}
P_R = \frac{3\lambda (1-\Phi^2)}{32}%, \ \ \ V=\frac{4}{3}\pi r_h^3. 
= \displaystyle \frac{3 \lambda \left(r_{h}^{2}-Q^{2}\right)}{32r_{h}^{2}}.
\end{eqnarray}
Within this framework, the conjugate quantity corresponding to the pressure is the thermodynamic volume, given by $V=\frac{4}{3}\pi r_h^3$\cite{Barzi:2023mit}. By integrating these elements, we can express the first law of Rényi thermodynamics and its associated Smarr formula as follows
\begin{eqnarray}
dM = T_RdS_R + VdP_R + \Phi dQ 
\qquad \text{ and }\qquad
M = 2T_RS_R - 2P_RV + \Phi Q. \label{Smarr_rényi_mod}
\end{eqnarray}
Likewise, we also define the specific volume 
\begin{equation}
v=\frac{8l_p^2}{3}r_h=\frac{8}{3}r_h,
\end{equation}
here $l_p$ denotes Planck's length, which has been set to unity for simplicity. Then the equation of state of the black hole thermodynamic fluid can be expressed such as  \cite{Barzi:2023mit}
\begin{equation}\label{bhr1}
P_R=  \frac{T_R}{v}  - \frac{2}{3 \pi v^{2}} + \frac{128 Q^{2}}{27 \pi v^{4}},
\end{equation}
Illustrating in Fig.\ref{fig:prenyiv} the equation of state given by Eq.\eqref{bhr1} unveils the presence of a critical behavior in the $P_R-v$ diagram, reminiscent of the behavior observed in the Van der Waals liquid-gas system. 
 
The black hole fluid exhibits a secondary phase transition at a critical temperature $T_{cr}=\displaystyle\frac{\sqrt{6}}{18 \pi Q}$, determined by solving the following system
\begin{equation}
	\left(\frac{\partial P_R}{\partial v}\right)_{T_{cr},Q}=
	\left(\frac{\partial^2 P_R}{\partial v^2}\right)_{T_{cr},Q}=0.
	\end{equation}

	For more details, we sketched a subcritical isotherm, $T_R<T_{cr}$ in the right panel of Fig.\ref{fig:prenyiv} and from which various features manifest:
 \begin{itemize}
     \item $v \; \in\; ]0,v_{min}[$ : $\partial_v P_R<0$, this domain corresponds to the stable (SBH) liquid-like phase.
     \item $v \; \in\; ]v_{max},\infty[ $ : $\partial_v P_R<0$, this is the (LBH) gas-like stable phase domain.
     \item $v \;\in\; ]v_{min},v_{max}[$ : $\partial_v P_R>0$, an unstable domain called \textit{the spinodal region}.
     \item $v=v_0$ : $\partial_{vv} P_R=0$, an inflection point within the spinodal region is observed. 
     \item $v \;\in \{v_{min},v_{max}\}$ : $\partial_v P_R=0$, two extrema points of the isotherm persist. 
 \end{itemize}

	In the next section, we will employ the Mel'nikov method to demonstrate the presence of either temporal chaos within the spinodal region or spatial chaos in the equilibrium configuration resembling Van der Waals-like black holes within the Rényi formalism.

\section{Chaos in the charged-flat black hole flow under thermal perturbations}
We adopt the model \cite{Chabab:2018lzf}, that the Rényi charged-flat black hole fluid is constrained to flow along the $x$-axis within a tube of fixed volume and a cross-section of one unit. Consequently, the position of any fluid particle is denoted by the coordinate $x$. The mass of a segment of the black hole fluid, which has a unit cross-section and extends from a reference particle at $x=0$ to any given fluid particle at position $x$, is calculated as follows
\begin{equation}\label{M_x_t}
m(x,t)=\int_{0}^{x}\rho(\tilde{x},t)d\tilde{x},
\end{equation}
where, $\rho(x,t)$ represents the density at position $x$ and time $t$, such a density is simply the inverse of the specific volume. Based on this definition, the position of any particle can be alternatively described in terms of mass by defining $x=x(m,t)$. Furthermore, leveraging this relationship enables the derivation that $x_m(m,t)\rho(x(m,t),t)=1$, where the subscript denotes a partial derivative, i.e., $x_m(m,t) \equiv \frac{\partial \displaystyle x}{\partial m}$. Thus, we obtain $x_m(m,t)= \big(\rho(x(m,t),t)\big)^{-1}$, which corresponds to the specific volume $v(m,t)$. Additionally, the velocity of the fluid can be defined as $u(m,t) \equiv \displaystyle \frac{\partial x}{\partial t}(m,t)$.

By imposing the conservation of mass and momentum, the flow of the Rényi charged-flat black hole can be effectively described through a mathematical model that takes the form of a dynamical system, as follows
\begin{align}
\partial_t v&=\partial_m u\label{conserv_eq1},\\
\partial_t u&=\partial_m \tau\label{conserv_eq2}.
\end{align}
Where, $\tau$ is the so-called Piola stress tensor\cite{felder:1970pio}. Considering a thermoelastic, isotropic, and slightly viscous black hole fluid in the absence of body forces, the $\tau$ tensor is given by 
\begin{equation}\label{piola}
\tau=\displaystyle-P(v,T)+\mu \partial_m u-A\partial_{mm}v.
\end{equation}
In this context, $A$ represents a positive constant, while $\mu$ is a small, positive constant representing viscosity. Combining Eqs. \eqref{conserv_eq1} and \eqref{conserv_eq2} with the substitution of Eq.\eqref{piola} we get
\begin{equation}\label{master_eq1}
\partial_{tt}x=\displaystyle-\partial_m P(v,T)+\mu \partial_{mm} u-A\partial_{mmm}v.
\end{equation}

%To simplify the mathematical treatment we introduce as in \cite{Dai_Chen_Jing:2020dil} a dimensionless parameter $s\in]0,1]$ such as the total mass of the black hole fluid verifies the condition $s=2\pi/M_{tot}$ and then by scaling all quantities with $s$ such as $m\longrightarrow sM$, $x\longrightarrow sx$ and $t\longrightarrow st$, we get $m\in[0,2\pi]$. We redefine the viscosity by, $\mu=\epsilon \mu_0$, where $\mu_0$ is a reference viscosity and $\epsilon$ measures the perturbation of the system. Thus Eq.\eqref{master_eq1} becomes
To facilitate the mathematical analysis, following the approach in \cite{Dai:2020wny}, we introduce a dimensionless parameter $s$ within the range $]0, 1]$ such that the total mass of the black hole fluid satisfies $s = 2\pi/M$. Subsequently, all relevant quantities are scaled by $s$: $m \rightarrow sm$, $x \rightarrow sx$, and $t \rightarrow st$, which normalizes $m$ to the interval $[0, 2\pi]$. Additionally, we redefine the viscosity as $\mu = \epsilon \mu_0$, where $\mu_0$ is a reference viscosity value and $\epsilon$ represents the perturbation scale. Consequently, the equation \eqref{master_eq1} becomes 
\begin{equation}\label{master_eq2}
\partial_{tt}x=\displaystyle-\partial_m P(v,T)-s\mu\; \partial_{mm} u-s^2A\;\partial_{mmm}v.
\end{equation}

Besides, in the vicinity  of the inflection point, the functions $ v(m, t) $ and $ u(m, t) $ can be expanded in Fourier series on the domain $ m \in[0, 2\pi ] $ respectively, 
%i.e., [21,22,29–31]
\begin{align}\label{v_and_u_fourier}
&v(m,t)=v_0+\sum_{k=1}^{\infty}\bar{v}_k(t)\cos(km),\\
&u(m,t)=\sum_{k=1}^{\infty}\bar{u}_k(t)\sin(km).
\end{align}
In this expansion, $k$ denotes the various oscillation modes induced in the phase space of the black hole fluid. For simplicity, we focus on the first modes, $(\bar{v}_1(t), \bar{u}_1(t))$, and for ease of notation, we will drop the subscript "1" henceforth. The equations are then expressed as follows
\begin{align}
&v(m,t)=v_0+\bar{v}(t)\cos(m)\label{v_one_mode},\\
&u(m,t)=\bar{u}(t)\sin(m)\label{u_one_mode}.
\end{align}
Now that we have defined the key equations that govern the system's dynamics, we will shift our focus to analyzing how thermal perturbations influence such dynamics.

\subsection{Chaos with time-dependent thermal perturbations}

Let us now explore the temporal chaos within the spinodal region where the phases of small and large black holes coexist. Initially, the system is in a stable homogeneous supercritical state characterized by $T_R > T_{cr}$ and a specific volume $v = v_0$. Subsequently, it is abruptly cooled to a subcritical temperature, $T_0 < T_{cr}$. This transition shifts the homogeneous state into the unstable spinodal region, maintaining the specific volume at $v = v_0$ (the inflection point), and sets the temperature to $T_R = T_0 < T_{cr}$. We proceed to analyze the dynamics under a small, time-periodic thermal perturbation around $T_0$ with the following form
\begin{equation}\label{T_peturb}
T=T_0+\epsilon \delta \cos(\omega t)\cos(m).
\end{equation}
Where $\delta$ represents the amplitude of the perturbation and $\omega$ its pulsation. We investigate their impact on the stability of the initially homogeneous and stable phase, potentially leading to the onset of temporal chaos. The expression for the Rényi pressure is expanded around the inflection point $(v_0, T_0)$ as follows
\begin{equation}\begin{split}
P(v_0,T)&=\displaystyle P_{0}+ P_{v} \left(v - v_{0}\right)+ P_{T} \left(T - T_{0}\right) + P_{Tv} \left(T - T_{0}\right) \left(v - v_{0}\right) + \frac{1}{2} P_{TT} \left(T - T_{0}\right)^{2}  \\ &+\frac{1}{2}P_{Tvv} \left(T - T_{0}\right) \left(v - v_{0}\right)^{2}  + \frac{1}{6}P_{vvv} \left(v - v_{0}\right)^{3} +\mathcal{O}((v-v_0)^4(T-T_0)^3).\end{split}
\label{pressure_expansion}
\end{equation}
Within this expansion, the coefficients are detailed in Table \ref{tab1}.

\begin{figure}[!ht]
	\centering
	\includegraphics[scale=0.6]{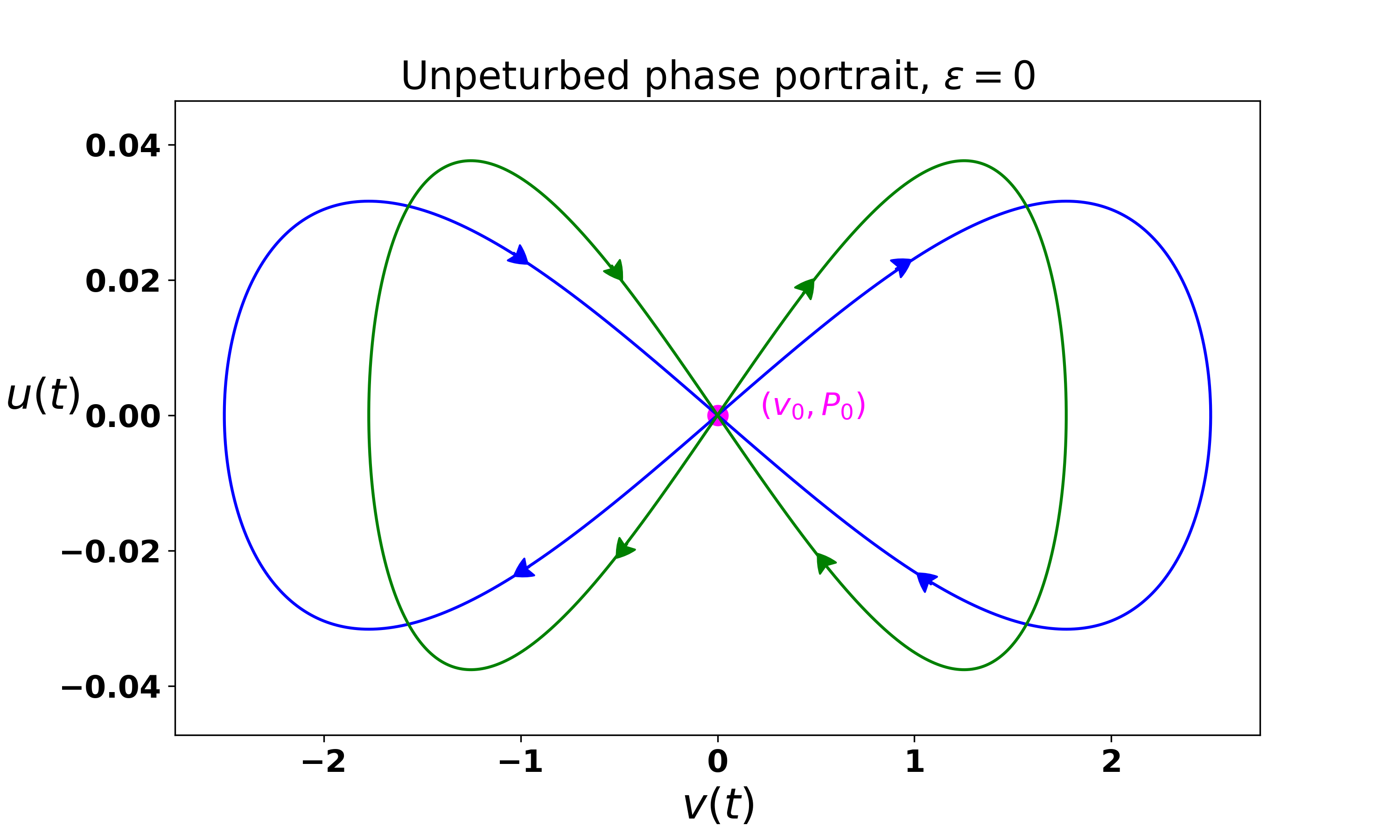}
	\caption{\footnotesize\it Homoclinic orbits of the unperturbed equations for the Rényi charged-flat black hole with $ T_0 = 0.5T_{cr}  $. The arrows indicate the flow of time. We set $Q =1/\sqrt{2} $ (green orbit), $Q=1. $ (blue orbit), $A = 0.01 $ and $ s = 0.001 $.}
	\label{fig:phase-portraitunpeturbed}
\end{figure}

\vspace{0.3cm}
\hspace*{-1cm}\begin{tabular}{||c||c||c||c||}
\hline 
	\rule[0ex]{0pt}{5ex}	$ P_0=\displaystyle \frac{128 Q^{2}}{27 \pi v_{0}^{4}} + \frac{T_{0}}{v_{0}} - \frac{2}{3 \pi v_{0}^{2}} $ & $ P_v=\displaystyle - \frac{512 Q^{2}}{27 \pi v_{0}^{5}} - \frac{T_{0}}{v_{0}^{2}} + \frac{4}{3 \pi v_{0}^{3}} $ & $ P_T=\displaystyle\frac{1}{v_0} $ & $ P_{Tv}=\displaystyle - \frac{1}{v_{0}^{2}} $ \\[2ex]
	\hline\hline
	\rule[0ex]{0pt}{5ex}$ P_{TT}=0 $ & $ P_{Tv}=\displaystyle - \frac{1}{v_{0}^{2}} $ & $ P_{Tvv}=\displaystyle \frac{2}{v_{0}^{3}} $ & 	$ P_{vvv}=\displaystyle - \frac{5120 Q^{2}}{9 \pi v_{0}^{7}} - \frac{6 T_{0}}{v_{0}^{4}} $ \\[2ex]
	\hline
	
\end{tabular}\captionof{table}{\footnotesize\it  Expansion coefficients of the Rényi pressure around the inflection point $(v_0,P_0,T_0)$.}\label{tab1}
\vspace{0.4cm}
 Notably, the coefficient $P_{vv}$ is zero, as $v = v_0$ represents an inflection point. Furthermore, $P_{vvv}$ is negative, reflecting a decrease in curvature from positive to negative values across the inflection point.
In this sense, the dynamical equations Eq.\eqref{conserv_eq1} and Eq.\eqref{conserv_eq2} become, using Eqs.\eqref{v_one_mode} and \eqref{u_one_mode},
\begin{equation}\label{eq_dynam_1}
\frac{d}{d t} \bar{v}{\left(t \right)}=\bar{u}(t).
\end{equation}
and
\begin{equation}\label{eq_dynam_2}
\begin{split}
\frac{d}{d t} \bar{u}{\left(t \right)}&=\displaystyle - A s^{2} \bar{v}{\left(t \right)} + P_{TT} \epsilon^{2} \delta^{2}  \cos{\left(m \right)} \cos^{2}{\left(\omega t \right)} + \frac{P_{vvv} \bar{v}^{3}{\left(t \right)} \cos^{2}{\left(m \right)}}{2} + P_{v} \bar{v}{\left(t \right)} \\ &+ \epsilon \delta  \left(\frac{3 P_{Tvv} \bar{v}^{2}{\left(t \right)} \cos^{2}{\left(m \right)}}{2} + 2 P_{Tv} \bar{v}{\left(t \right)} \cos{\left(m \right)} + P_{T}\right) \cos{\left(\omega t \right)} - \epsilon \mu_{0} s \bar{u}{\left(t \right)}.
\end{split}
\end{equation}
For small values of $\epsilon$, where $\epsilon<<1$, the $\epsilon^2$-term can be neglected. It is important to note that the equations retain a dependence on the scaled mass $m$, indicating that fluid particles are not equivalent. As a result, different columns of fluid will respond differently to time-dependent thermal perturbations. For the unperturbed black hole scenario, by setting $\epsilon=0$, we arrive at the following dynamical system
\begin{eqnarray}\label{unpertb_dynm_22X}
\frac{d}{d t} \bar{v}{\left(t \right)}&=&\bar{u}(t),\\
\frac{d}{d t} \bar{u}{\left(t \right)}&=&\displaystyle \frac{P_{vvv} \bar{v}^{3}{\left(t \right)} \cos^{2}{\left(m \right)}}{2} + \left(- A s^{2} + P_{v}\right) \bar{v}{\left(t \right)}.\label{unpertb_dynm_2}
\end{eqnarray}
By incorporating $\theta(t)=\begin{bmatrix}\bar{v}(t)\\\bar{u}(t)\end{bmatrix}$, we can rewrite the above systems of equations in a more compact form such as
\begin{equation}\label{temp_chaos_petrb}
\frac{d}{d t} \theta{\left(t \right)}=f(\theta)+\epsilon h(\theta,t),
\end{equation}
Where the functions $f$ and $h$ are defined by,
\begin{equation}\label{f_time}
f(\theta)=
\begin{bmatrix} 0\\ \displaystyle \frac{P_{vvv} \bar{v}^{3}{\left(t \right)} \cos^{2}{\left(m \right)}}{2} + \left(- A s^{2} + P_{v}\right) \bar{v}{\left(t \right)},
\end{bmatrix}
\end{equation}
and
\begin{equation}\label{h_time}
h(\theta,t)=
\begin{bmatrix} 0\\ \displaystyle \frac{ \delta\left[3 P_{Tvv} \bar{v}^{2}{\left(t \right)} \cos^{2}{\left(m \right)} + 4 P_{Tv} \bar{v}{\left(t \right)} \cos{\left(m \right)} + 2 P_{T}\right] \cos{\left(\omega t \right)}}{2} - \mu_{0} s \bar{u}{\left(t \right)}
\end{bmatrix}.
\end{equation}
To utilize the Mel'nikov method, we must first solve the unperturbed equation $\frac{d}{dt}\theta(t)=f(\theta)$. Indeed,
the unperturbed black hole fluid admits the following solutions \cite{holmes1979:one}
\begin{equation}\label{unpert_sol}
\theta_0(t)=
\begin{bmatrix}
\displaystyle \pm\frac{2 w}{\sqrt{- P_{vvv}} \cosh{\left( w t\right)} \left|{\cos{\left(m \right)}}\right|}\\ \displaystyle  \mp\frac{2 w^{2} \sinh{\left( w t\right)}}{\sqrt{- P_{vvv}} \cosh^{2}{\left(w t\right)} \left|{\cos{\left(m \right)}}\right|}
\end{bmatrix} .
\end{equation}
Where $w^2=\displaystyle - A s^{2} + P_{v}$. 
The solution Eq.\eqref{unpert_sol} describes a {\it homoclinic} orbit that connects a saddle equilibrium point back to itself. It consists of two branches, each corresponding to a different sign choice, together forming the butterfly-shaped orbit illustrated in Fig.\ref{fig:phase-portraitunpeturbed}.
We note that in the absence of a perturbation, the two branches are independent of each other, such that if the system starts on one of the branches it stays on it permanently.

\begin{figure}[ht!]

			\begin{tabbing}
        \hspace{-1.cm}
			%\hspace{9.3cm}
   %\=\kill
   \includegraphics[width=9.5cm,height=7cm]{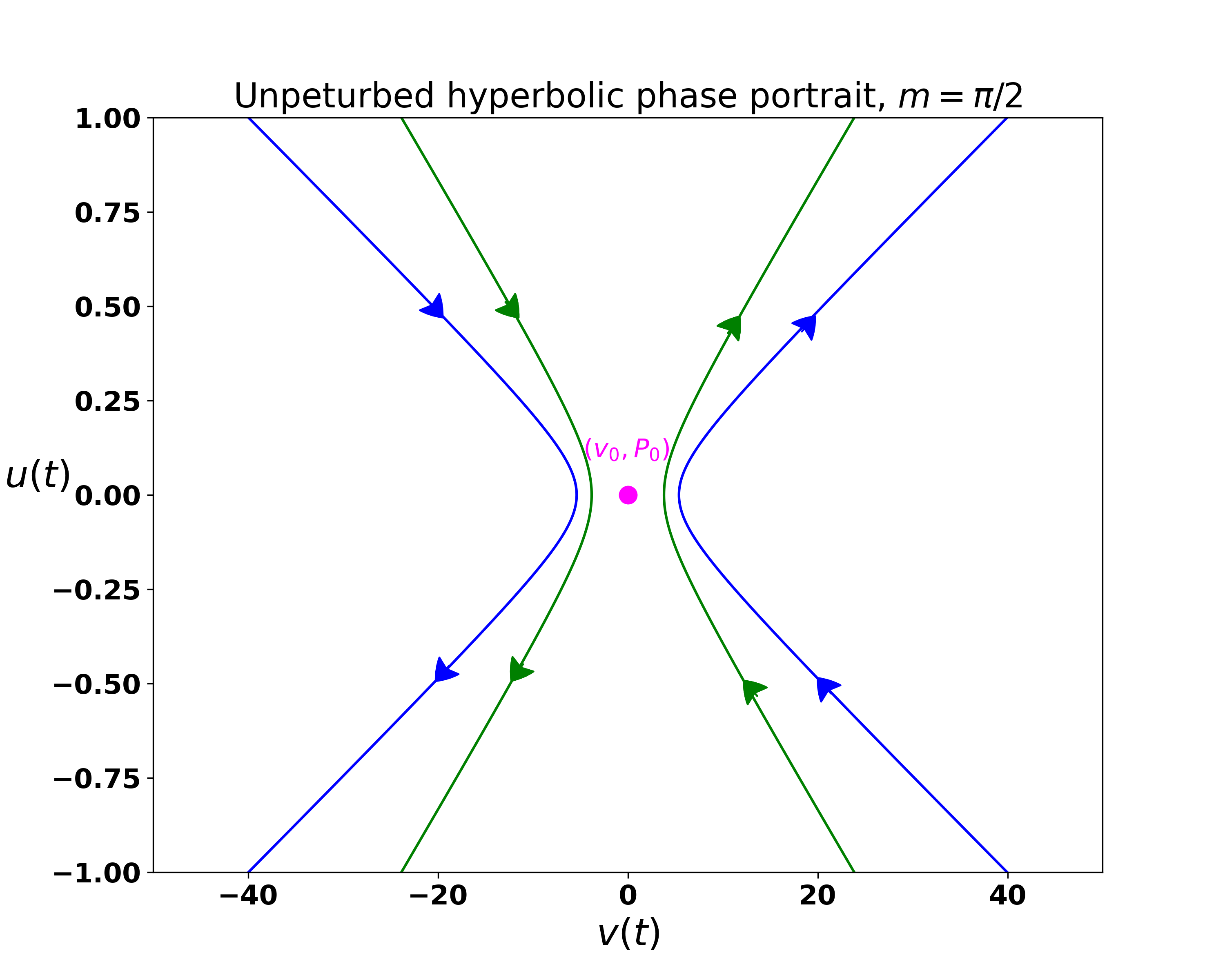}
  %\>
		\hspace{-1.cm}
  \includegraphics[width=9.5cm,height=7cm]{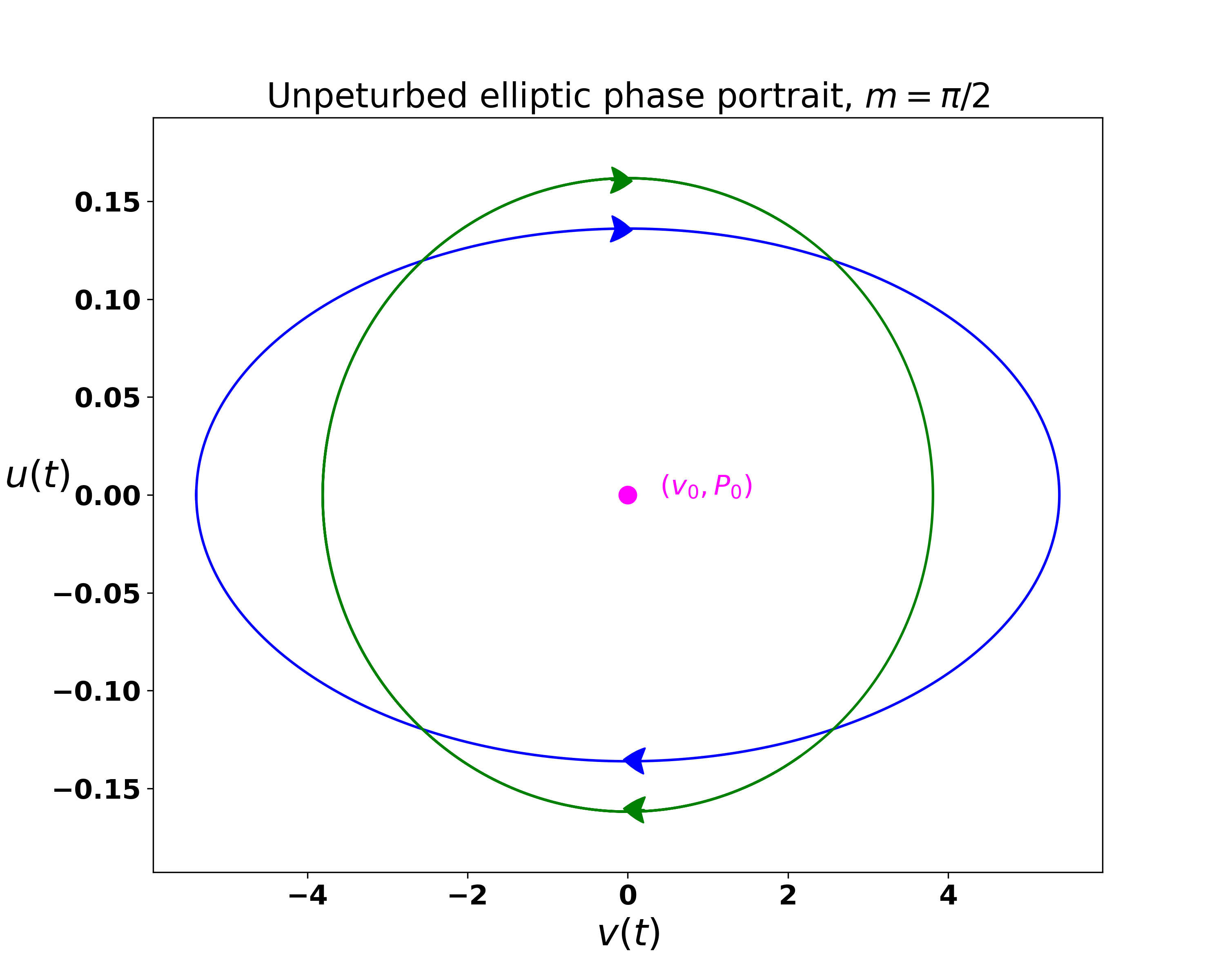}
	\end{tabbing}\vspace{-.8cm}
	\caption{\footnotesize\it {\bf(Left:)} Hyperbolic and {\bf(Right:)} Elliptic orbits of the unperturbed equations for the Rényi charged-flat black hole with $T_0 = 0.5T_{cr}$ and scaled fluid mass $m=\pi/2$. The arrows indicate the flow of time. We set $Q =1/\sqrt{2} $ (green orbit), $Q=1. $ (blue orbit), $A = 0.01 $ and $ s = 0.001 $.}
	\label{fig:phase-portraitunpeturbed_M_half_pi}
\end{figure}

\paragraph{} However, there is a notable exception for fluid columns with a scaled mass of $m=\frac{\pi}{2}$(or $\frac{3\pi}{2}$), which are not accounted for in the general solution provided by Eq.\eqref{unpert_sol}. In these cases, the nonlinear term in Eq.\eqref{unpertb_dynm_2} vanishes, simplifying the dynamics of the system as described by the following equation
\begin{align}
\frac{d^2}{d t^2} \bar{v}{\left(t \right)}&=\displaystyle \left(- A s^{2} + P_{v}\right) \bar{v}{\left(t \right)},\label{unpertb_dynm_2_lin}
\end{align}
Which is a linear equation easily solved such as
\begin{equation}\label{unpert_sol_pi_over_2_case1}
\theta_{0,m=\frac{\pi}{2}}(t)=
\begin{bmatrix}
\displaystyle v_0\cos(wt)\\  -wv_0 \sin(wt)
\end{bmatrix} \quad for \quad w^2<0,
\end{equation}
and, 
\begin{equation}\label{unpert_sol_pi_over_2_case2}
\theta_{0,m=\frac{\pi}{2}}(t)=
\begin{bmatrix}
\displaystyle v_0\cosh(wt)\\  -wv_0 \sinh(wt)
\end{bmatrix} \quad for\quad w^2>0.
\end{equation}
These special solutions, Eqs.\eqref{unpert_sol_pi_over_2_case1} and \eqref{unpert_sol_pi_over_2_case2}, correspond to {\it elliptic} and {\it hyperbolic} orbits respectively as depicted in Fig.\ref{fig:phase-portraitunpeturbed_M_half_pi}.

These findings apply equally to the unperturbed charged-AdS black hole fluid, where the pressure is associated to the negative cosmological constant, $\Lambda<0$, such as, $P=-\frac{\Lambda}{8\pi}$, instead of Eq\eqref{p-lam}.

\subsection{Mel'nikov function for time-dependent perturbations}
\begin{figure}[ht!]
			\begin{tabbing}
       \hspace{-0.9cm}
			% \hspace{9.3cm}\=\kill
	\includegraphics[width=9.5cm,height=7.5cm]{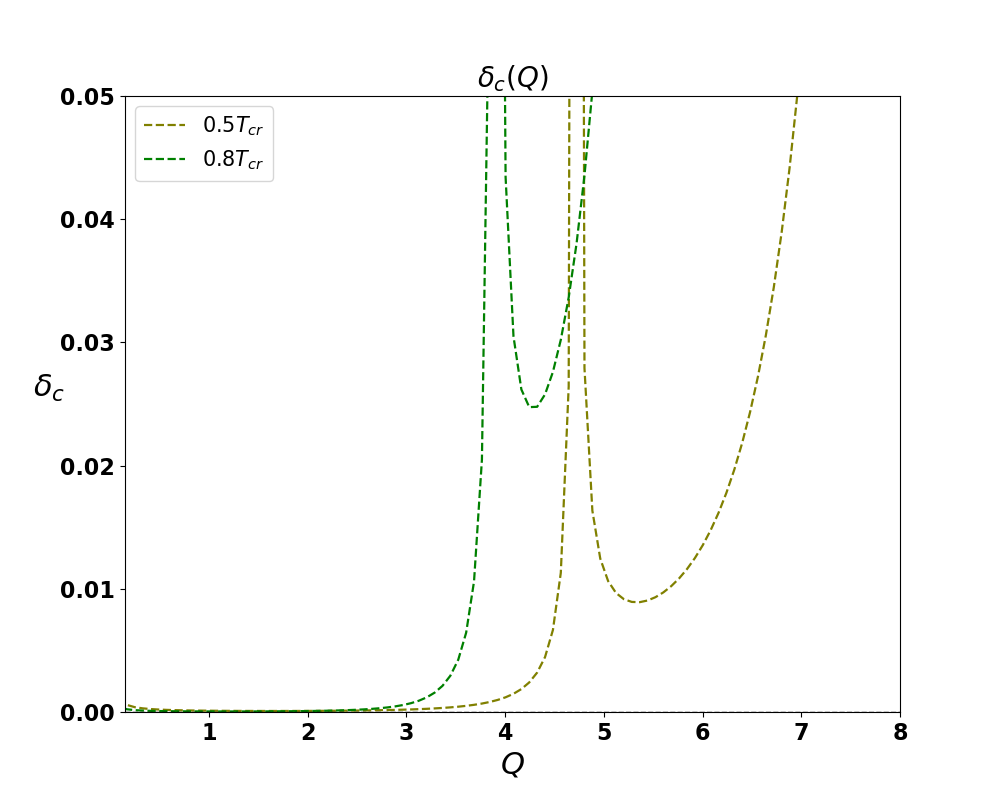}%\>
		\hspace{-0.7cm}
		\includegraphics[width=9.5cm,height=7.5cm]{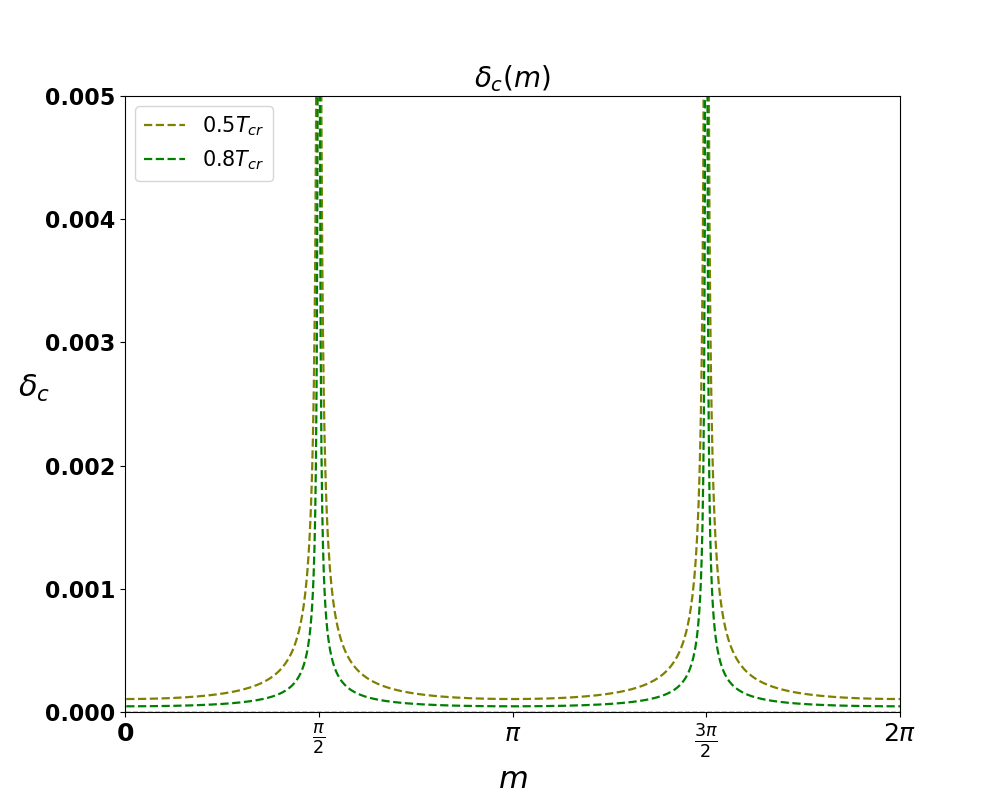}\\
   \hspace{3cm} 
	\includegraphics[width=9.5cm,height=7.5cm]{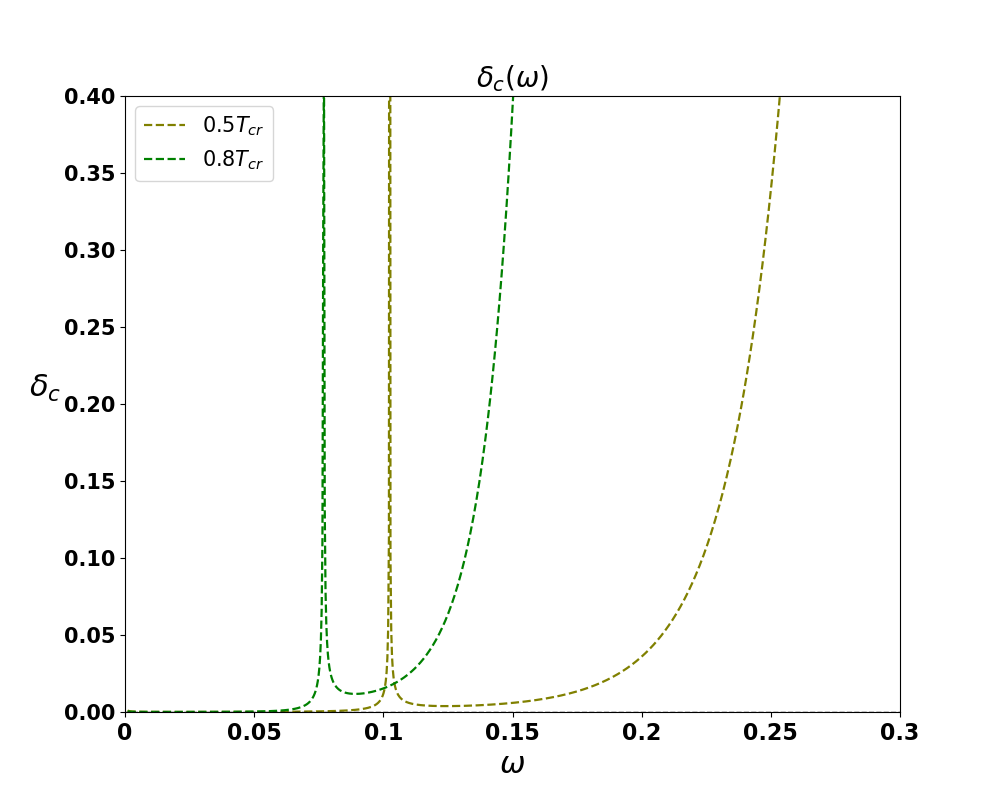}
	\end{tabbing}
 \vspace{-1.cm}
	\caption{\footnotesize\it Variations of the critical value $\delta_c$ with {\bf (Top left:)}  the black hole electric charge $Q$, {\bf(Top right:)} The scaled mass $m$ of the black hole fluid column and {\bf (Bottom:)} The pulsation $\omega$ of the periodic temporal perturbation . Other parameters are fixed as $\omega=0.01$(top left and right), $m=2\pi$(top left and bottom), $Q=1.$(top right and bottom), $A = 0.01 $ and $ s = 0.001 $.}
	\label{fig:delta_c}
\end{figure}

The existence of chaos is examined through the Mel'nikov function, which is defined for a given initial time $t_0$ by
\begin{equation}\label{Melnk_def}
\mathbf{\mathcal{M}}(t_0)=\int_{-\infty}^{\infty}f^T(\theta_0(t))\;\mathcal{\bm J}\;h(\theta_0(t),t)\bigg\rvert_{t=\tilde{t}-t_0}d\tilde{t}.
\end{equation}
Where we define the anti-symmetric matrix $\mathcal{J}=\displaystyle \left[\begin{matrix}0 & 1\\-1 & 0\end{matrix}\right]$. 
 The Mel'nikov function $\mathcal{M}(t_0)$, serves as an estimate of the separation between the stable and unstable manifolds in the phase space. If $\mathcal{M}(t_0)$ exhibits a simple zero at $t_0$, then, for a sufficiently small $\epsilon>0$, the stable and unstable manifolds of the Hamiltonian system will intersect transversally~\cite{Guckenheimer1983NonlinearOD,Chabab:2018lzf,doi:10.1098/rsta.1979.0068,HOLMES1990137,HOLMES40}. According to the Smale-Birkhoff homoclinic theorem~\cite{Birkhoff,SmaleS42}, such transversal intersections indicate the presence of an invariant hyperbolic set, known as a Smale horseshoe, in the Poincaré map. This configuration is a strong indicator of chaotic dynamics in the system~\cite{HOLMES1990137}. The existence of a little viscous contribution to the stress tensor Eq.\eqref{piola} is essential to our interpretation. This does not imply that the inviscid case won't have horseshoes; rather, it only indicates that the Mel'nikov approach employed here is unsuitable for inviscid flows.  
Injecting the Eq.\eqref{f_time} and Eq.\eqref{h_time} into Eq.\eqref{Melnk_def} and denoting $\cos(m)$ by $y$ $(m\neq \{\frac{\pi}{2},\frac{3\pi}{2}\})$ , we obtain the following integral
\begin{equation}\label{Mel_t0}
\begin{split}
\mathbf{\mathcal{M}}(t_0)&=\int_{-\infty}^{\infty}\displaystyle \frac{12 P_{Tvv}  w^{4} \delta\cos{\left(\omega t \right)} \sinh{\left[w(t - t_{0})  \right]}}{P_{vvv} y \sqrt{- P_{vvv}} \cosh^{4}{\left[w(t - t_{0}) \right]}}dt +\int_{-\infty}^{\infty} \frac{8 P_{Tv}  w^{3}\delta \cos{\left(\omega t \right)} \sinh{\left[w(t - t_{0}) \right]}}{P_{vvv} y \cosh^{3}{\left[w(t - t_{0}) \right]}}dt \\\\&-\int_{-\infty}^{\infty} \frac{2 P_{T}  w^{2}\delta \cos{\left(\omega t \right)} \sinh{\left[w(t - t_{0}) \right]}}{y \sqrt{- P_{vvv}} \cosh^{2}{\left[w(t - t_{0}) \right]}}dt +\int_{-\infty}^{\infty} \frac{4 \mu_{0} s w^{4} \sinh^{2}{\left[w(t - t_{0})  \right]}}{P_{vvv} y^{2} \cosh^{4}{\left[w(t - t_{0})\right]}}dt.
\end{split}
\end{equation}
Through a direct integration, term by term, we obtain the Mel'nikov function in the following compact form
\begin{equation}\label{key}
\mathbf{\mathcal{M}}(t_0)=\mathcal{L}+\mathcal{K}\delta \sin(\omega t_0).
\end{equation}
Where,
\begin{align}\label{key}
\mathcal{L}&=\displaystyle \frac{8 \mu_{0} s w^{3}}{3 P_{vvv} y^{2}},\\
\mathcal{K}&=\displaystyle \frac{2 \pi \omega}{y \left(- P_{vvv}\right)^{\frac{3}{2}}\cosh{\left(\frac{\pi \omega}{2 w} \right)}} \left\{ P_{Tvv}  \left(\omega^{2} + w^{2}\right)- \omega P_{vvv}P_{T} + \frac{2 \omega \sqrt{- P_{vvv}} P_{Tv}}{\tanh{\left(\frac{\pi \omega}{2 w} \right)}}  \right\}.\label{K}
%\displaystyle \frac{2 \pi P_{Tvv} \omega \left(\omega^{2} + w^{2}\right)}{y \left(- P_{vvv}\right)^{\frac{3}{2}} \cosh{\left(\frac{\pi \omega}{2 w} \right)}} - \frac{4 \pi P_{Tv} \omega^{2}}{P_{vvv} y \sinh{\left(\frac{\pi \omega}{2 w} \right)}} + \frac{2 \pi P_{T} \omega}{y \sqrt{- P_{vvv}} \cosh{\left(\frac{\pi \omega}{2 w} \right)}}
\end{align}
%The equation $\mathbf{\mathcal{M}}(t_0)=0$ possesses real solutions if $\delta$ satisfy the inequality $\displaystyle\delta \geq\biggr\rvert \frac{\mathcal{L}}{\mathcal{K}}\biggr\rvert$ while $\mathcal{K}\neq 0$. We pose $\displaystyle \delta_c =\biggr\rvert \frac{\mathcal{L}}{\mathcal{K}}\biggr\rvert$, which is the critical value for the onset of thermal chaos in the black hole fluid due to time-dependent periodic perturbations. 
The equation $\mathbf{\mathcal{M}}(t_0) = 0$ possesses real solutions when $\delta$ satisfies the inequality $\displaystyle \delta \geq \left| \frac{\mathcal{L}}{\mathcal{K}} \right|$, while $\mathcal{K} \neq 0$. We define $\displaystyle \delta_c = \left| \frac{\mathcal{L}}{\mathcal{K}} \right|$ as the critical value at which thermal chaos is initiated in the black hole fluid due to time-dependent periodic perturbations. The variation of the quantity $\delta_c$ with the electrical charge $Q$ and scaled mass $m$ is illustrated in Fig.\ref{fig:delta_c}. An additional computation yields the following formula for $\delta_c$,
\begin{equation}\label{delta_expression}
    \delta_c=\displaystyle  \frac{\mu_{0} s v_{0}^{2} w^{3}}{3 \pi \omega^2 y}\frac{ \displaystyle \sinh{\left(\frac{\pi \omega}{2w} \right)} }{ \left[\displaystyle 1   - \left(\frac{\omega^{2} + w^{2}+q^{2} v_{0}^{2}}{\omega q v_{0}}\right) \tanh{\left(\frac{\pi \omega}{2 w} \right)}\right]},
\end{equation}
here we have put $q^2=-P_{vvv}$. 

\paragraph{}The top left panel shows that $\delta_c$ approaches infinity at specific values of the electric charge $Q_r$ ($Q_r=4.7170$ at 0.5$T_{cr}$ and $Q_r=3.8961$ at 0.8$T_{cr}$), indicating that the black hole fluid becomes completely resistant to temporal thermal perturbations at these charges. This suggests that higher electric charges significantly suppress the potential for time-dependent chaos, highlighting the stabilizing influence of the electric charge. On the top right panel, we observe that the onset of chaos becomes increasingly difficult for fluid columns with a scaled mass around $\displaystyle\frac{\pi}{2}$ (and $\displaystyle\frac{3\pi}{2}$), indicating a higher resistance to chaotic conditions in these regions. On the bottom panel, we plotted the spectral response of the chaos threshold which presents a low-pass behavior permitting a soft response to slowly varying fluctuations but a hard response to rapidly changing ones. Moreover, a narrow anti-resonance is observed at particular pulsations $\omega_r$ ($\omega_r=0.1022$ at $0.5T_{cr}$ and $\omega_r=0.0767$ at $0.8T_{cr}$). This novel character points towards a potential connection with the quasinormal modes of the charged-flat black hole within R\'enyi formalism to be explored further .\\
 
\begin{figure}[!ht]
	\centering
 \begin{tabbing}
 %    \hspace{-2.7cm}
	\includegraphics[scale=0.4]{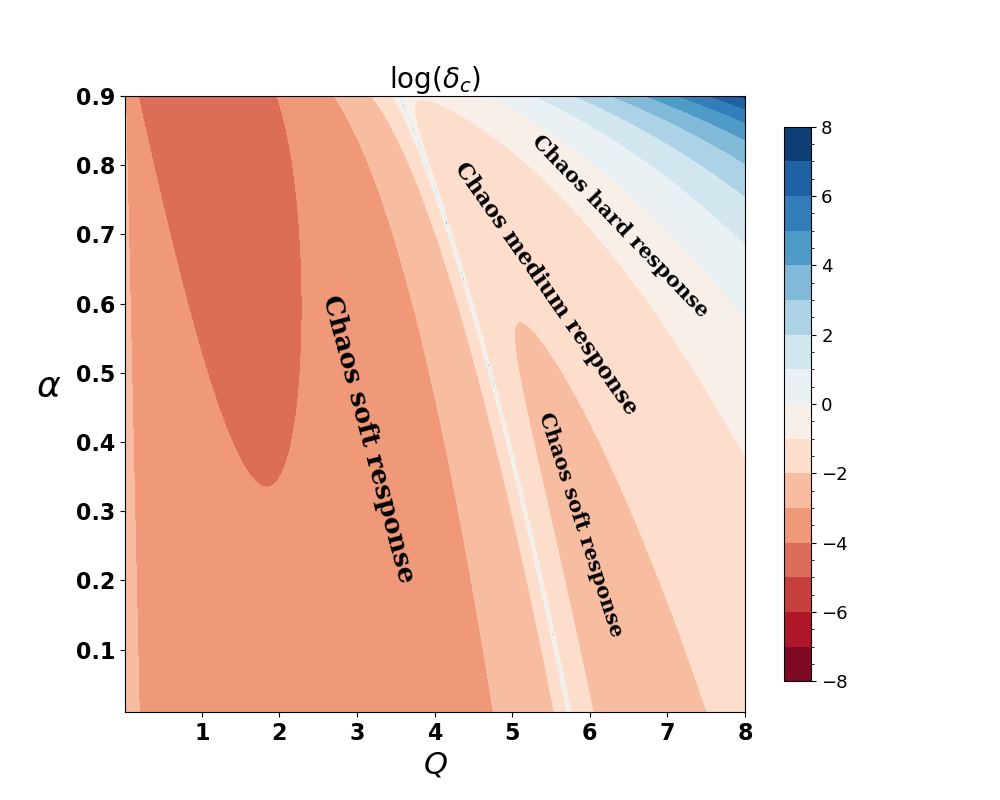}
 \hspace{-1.cm}
     \includegraphics[scale=0.4]{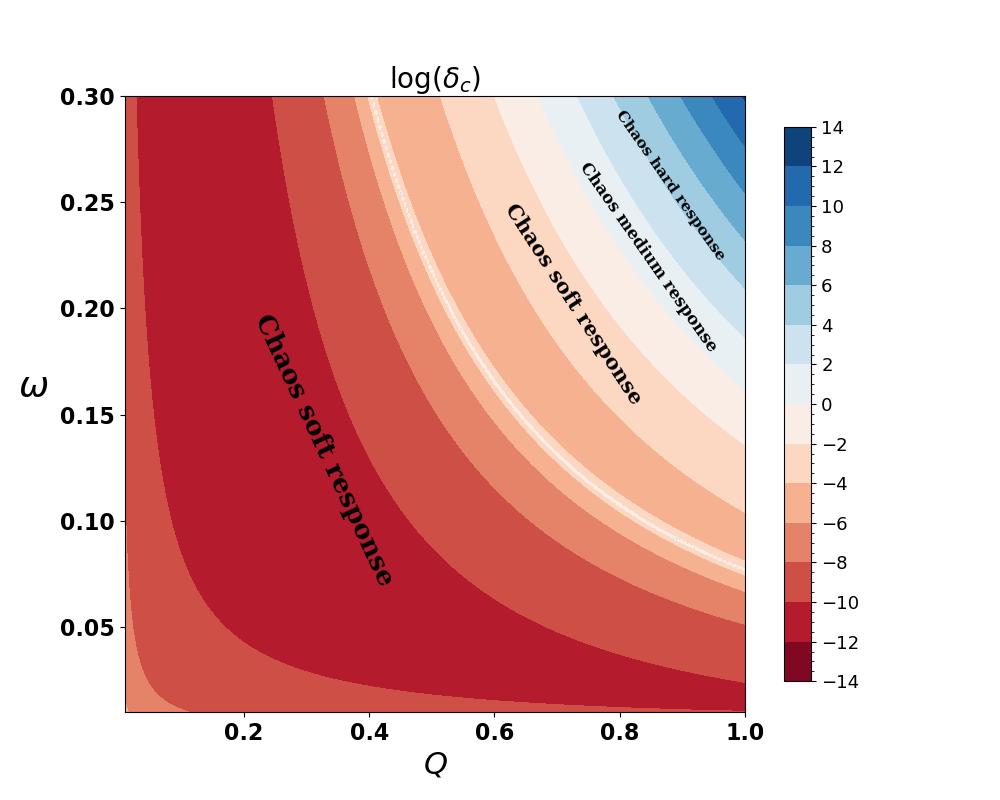}
 \end{tabbing}
	\vspace{-0.8cm}
 \caption{\footnotesize\it Maps in {\bf (Left:)} $(\alpha,Q)$-plan and {\bf(Right:)}  $(\omega,Q)$-plan depicting the evolution of the critical value $\delta_c$ with the electric charge $Q$, the ratio $\alpha=\frac{T_0}{T_{cr}}$ and the pulsation of time-periodic thermal  $\omega$. The plots unveil three regions of the chaotic scheme: chaos soft response, medium response, and hard response. Other parameters are fixed as $\omega=0.01$(left), $\alpha=0.8$(right), $m=2\pi$, $A = 0.01 $ and $ s = 0.001 $.}
	\label{fig:delta_c_map}
\end{figure}

In Fig.\ref{fig:delta_c_map}, we illustrate various chaotic regions within the semi-logarithmic $\alpha-Q$ and $\omega-Q$ parameter spaces. Here,  $\alpha$ is defined as $\alpha=\frac{T_0}{T_{cr}}$, indicating the ratio of the coexistence temperature $T_0$ to the critical temperature $T_{cr}$.

The depicted regions correspond to distinct chaotic behavior of the black hole fluid in response to time-periodic thermal fluctuations. On the left panel, the first identified region is labeled as the {\it chaos soft response region}. In this area, thermal chaos is readily induced with the fluid characterized by a relatively low electric charge, up to $Q \approx 4.75$, and a coexistence temperature up to $T_0 \approx 0.83\;T_{cr}$. Within this region, the chaos threshold remains low, with $\delta_c < 10^{-2}$. The intermediate region is a {\it chaos medium response} domain, where the fluid is less prone to thermal chaos with a threshold in the range, $10^{-2}<\delta_c<1$, notable also is a substantial domain of soft response in the range, $6.1<Q<7.5$, within this medium region. The third region is a {\it chaos hard response} domain where the black hole fluid is almost unaffected by time-dependent thermal perturbations and presents large critical values $\delta_c>1$. This domain is marked by a thin band separating the soft and medium regions, in addition to a domain of large electric charge, $Q>8$, and coexistence temperatures higher than $0.4\;T_{cr}$, which exhibits extreme values of the chaos threshold, $\delta_c>10^{2}$, confirming the stabilizing effect of the electric charge. This effect can plausibly be attributed to the repulsive nature of the Coulombian interaction, as it tends to restrict the motion of the fluid particles, keeping them at sufficient distances from each other and thereby limiting their collisions. The same regions are manifested on the right panel, here the soft response dominates at low electric charges but the hard response is prevailing as pulsations become larger, starting at $\omega>0.16$. Furthermore, a narrow strip of hard response is revealed within the soft domain signaling a anti-resonance at specific values of the the electric charge and pulsation. Consequently, the black hole fluid behavior is analogous to that of a low-pass filter and Eq.\eqref{delta_expression} can be interpreted as the {\it transfer function} of the R\'enyi charged-flat black hole in response to thermal time-periodic chaos. 

\begin{figure}[!ht]
\centering 
			\begin{tabbing}
   \hspace*{-1.7cm}
			\centering
			\hspace{10.cm}\=\kill
   \includegraphics[width=.54\textwidth]{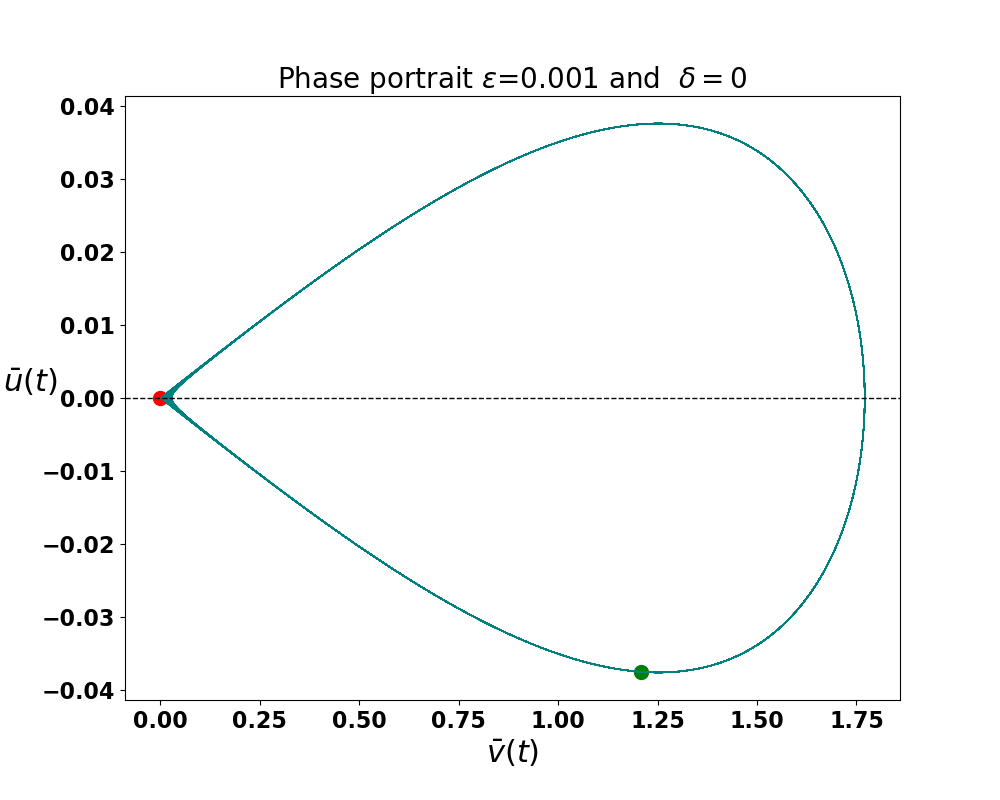}\>
\includegraphics[width=.54\textwidth]{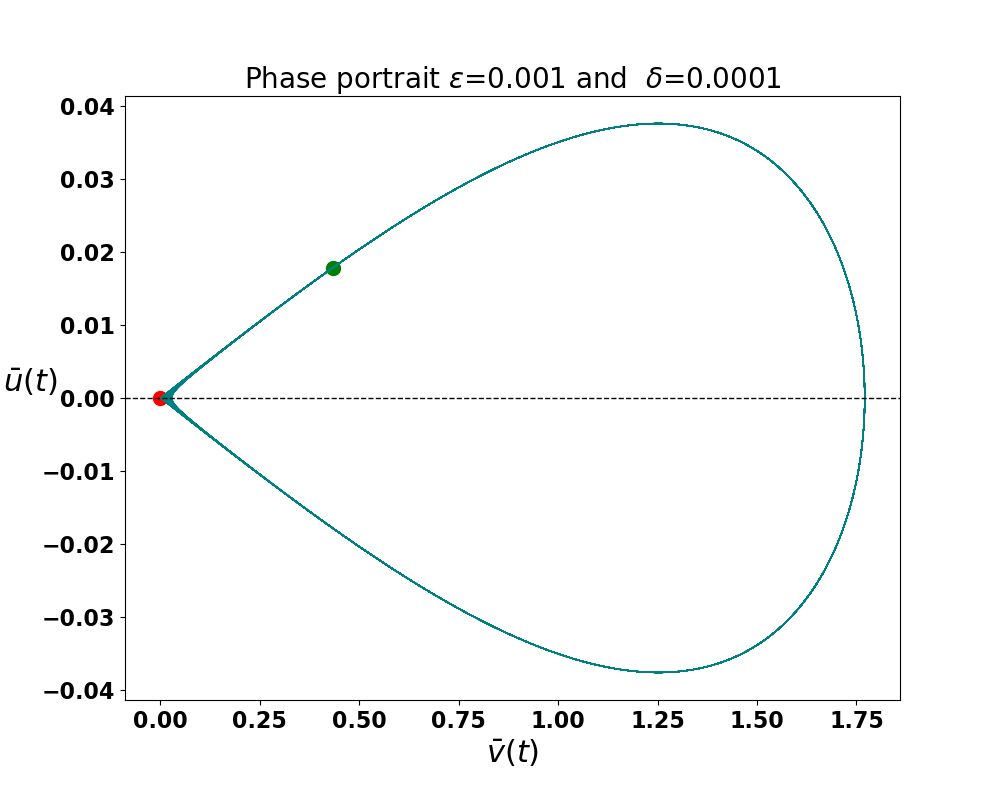}\\
   \includegraphics[width=.54\textwidth]{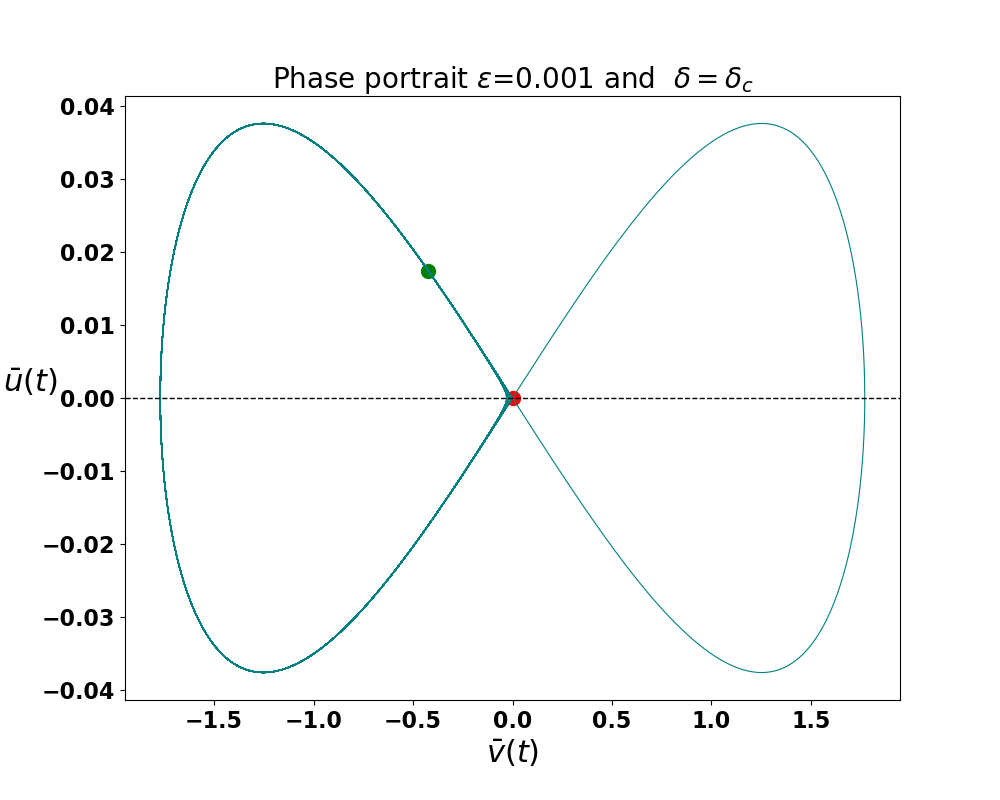}\>
\includegraphics[width=.54\textwidth]{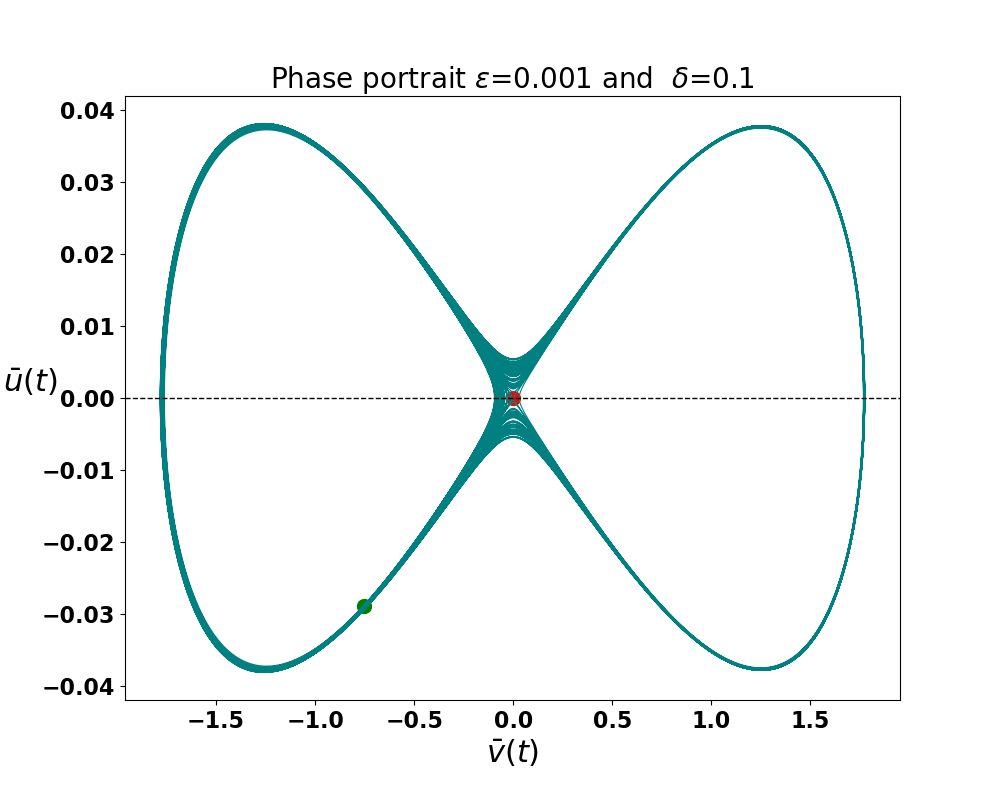}\\
\includegraphics[width=.54\textwidth]{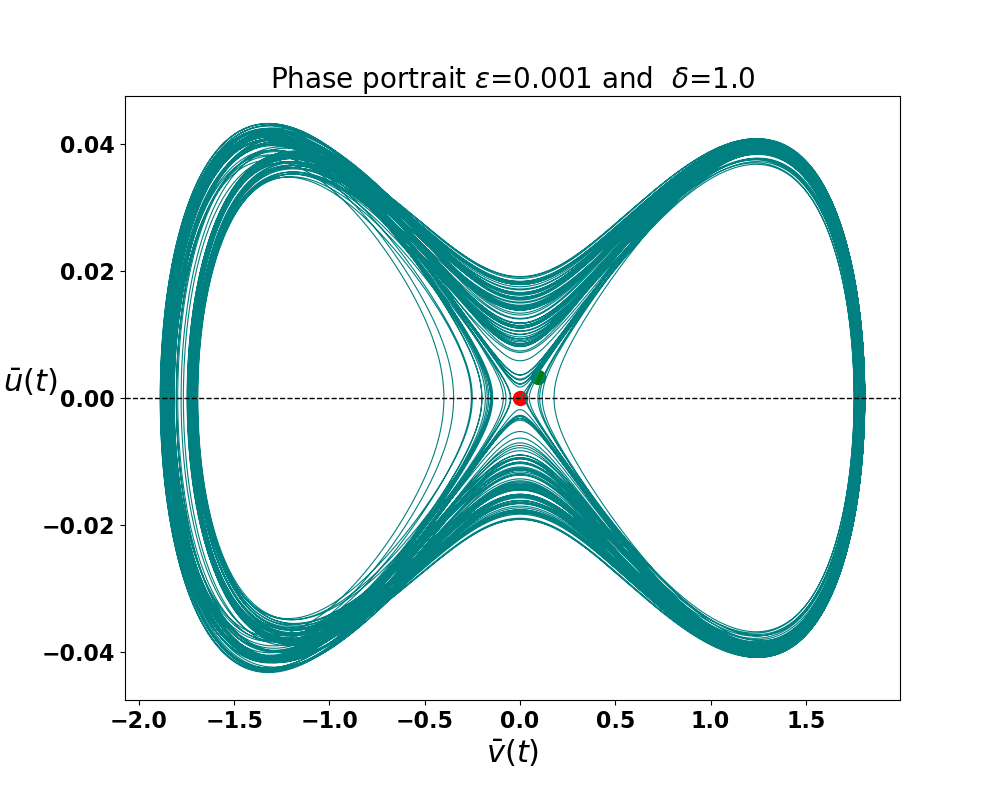}\>
\includegraphics[width=.54\textwidth]{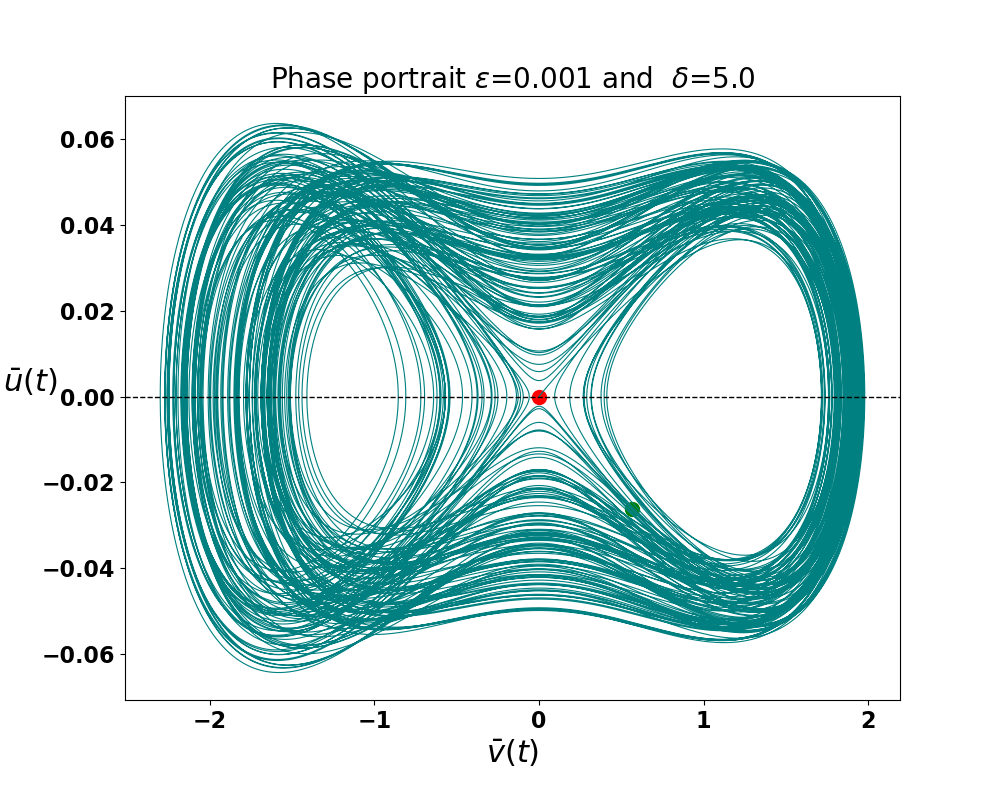}\\
\end{tabbing}\vspace{-.8cm}
\caption{ \footnotesize   \it
Evolution of the perturbed solutions with time $ t $ in $ \bar{u} - \bar{v} $ plane for the fixed system temperature $ T_R = 0.5 T_{cr} $. Parameters are fixed as $\omega = 0.01$, $\epsilon = 0.001$, $\mu_0 = 0.1$. The critical value is $\delta_c\approx 0.00014144$. The red and green dots denote the initial and final phases, respectively.
}
\label{fig:6}
\end{figure}

\paragraph{}Chaotic behavior is illustrated in Fig.\ref{fig:6}, where we present numerical simulations of the equations of motion Eqs.\eqref{temp_chaos_petrb}, \eqref{f_time} and \eqref{h_time}. The top plots in Fig.\ref{fig:6} depict typical system trajectories under normal conditions, without and with a small perturbation respectively, for $\delta < \delta_c$. The rest of the panels capture the emergence of chaotic trajectories when $\delta$ exceeds the critical threshold, $\delta_c$. These chaotic responses as well as the chaos limiting effect of the electric charge are present in the charged-AdS fluid as demonstrated in \cite{Chabab:2018lzf,Dai:2020wny}, which substantiates once more the conjectured correspondence between the R\'enyi charged-flat and the Gibbs-Boltzmann charged-AdS black holes. We turn our attention now to probing spatially initiated chaos within the coexistence region of the two black hole phases (SBH/LBH).

%\begin{figure}[!ht]
%\vspace{-1cm}\hspace{-10mm}
%	\subfloat[]{
%		\includegraphics[scale=0.28]{phase portrait_peturbed_3}
%	}\hspace{-9mm}
%	\subfloat[]{   
%		\includegraphics[scale=0.28]{phase portrait_peturbed_4}
%	}\hspace{-9mm}
%	\subfloat[]{
%		\includegraphics[scale=0.28]{phase portrait_peturbed_5}
%	}
%	\caption{Evolution of the perturbed equations with time $ t $ in $ x - u $ plane for the fixed system temperature $ T_R = 0.5 T_{cr} $. Parameters are fixed as $\omega = 0.01$, $\epsilon = 0.001$, $\mu_0 = 0.1$. The critical value $\delta_c\approx 0.00014144$. The red and green dots denote the initial and final phases, respectively.
%	}
%\end{figure}
%

\newpage 
\subsection{Chaos with space-dependent thermal perturbations}

\begin{figure}[ht!]
\vspace{-0.9cm}
	\begin{tabbing}
	\hspace{-1.5cm}
   % \=\kill
   \includegraphics[scale=0.4]{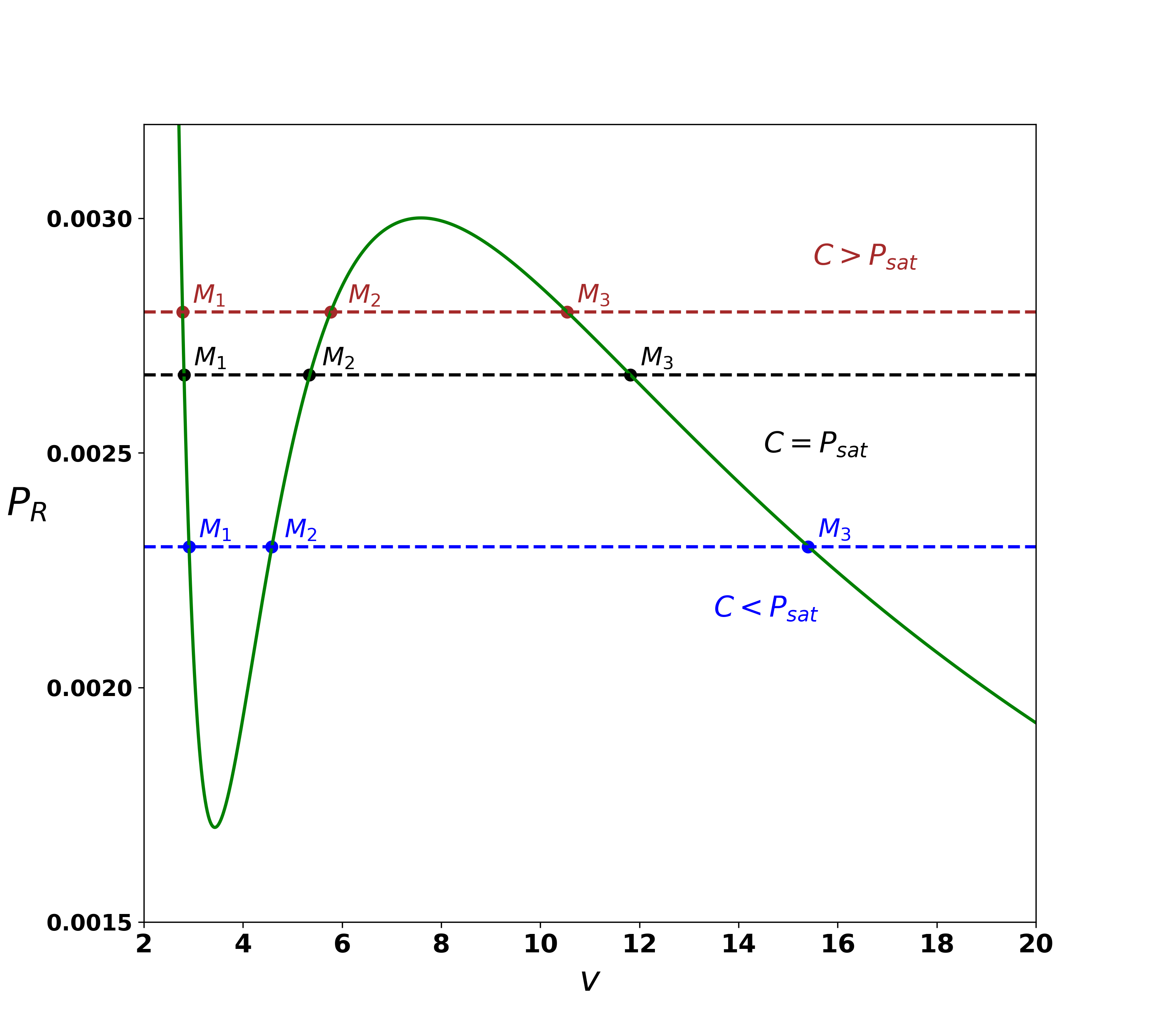}
   %\>
   	\hspace{-1cm}
	\includegraphics[scale=0.4]{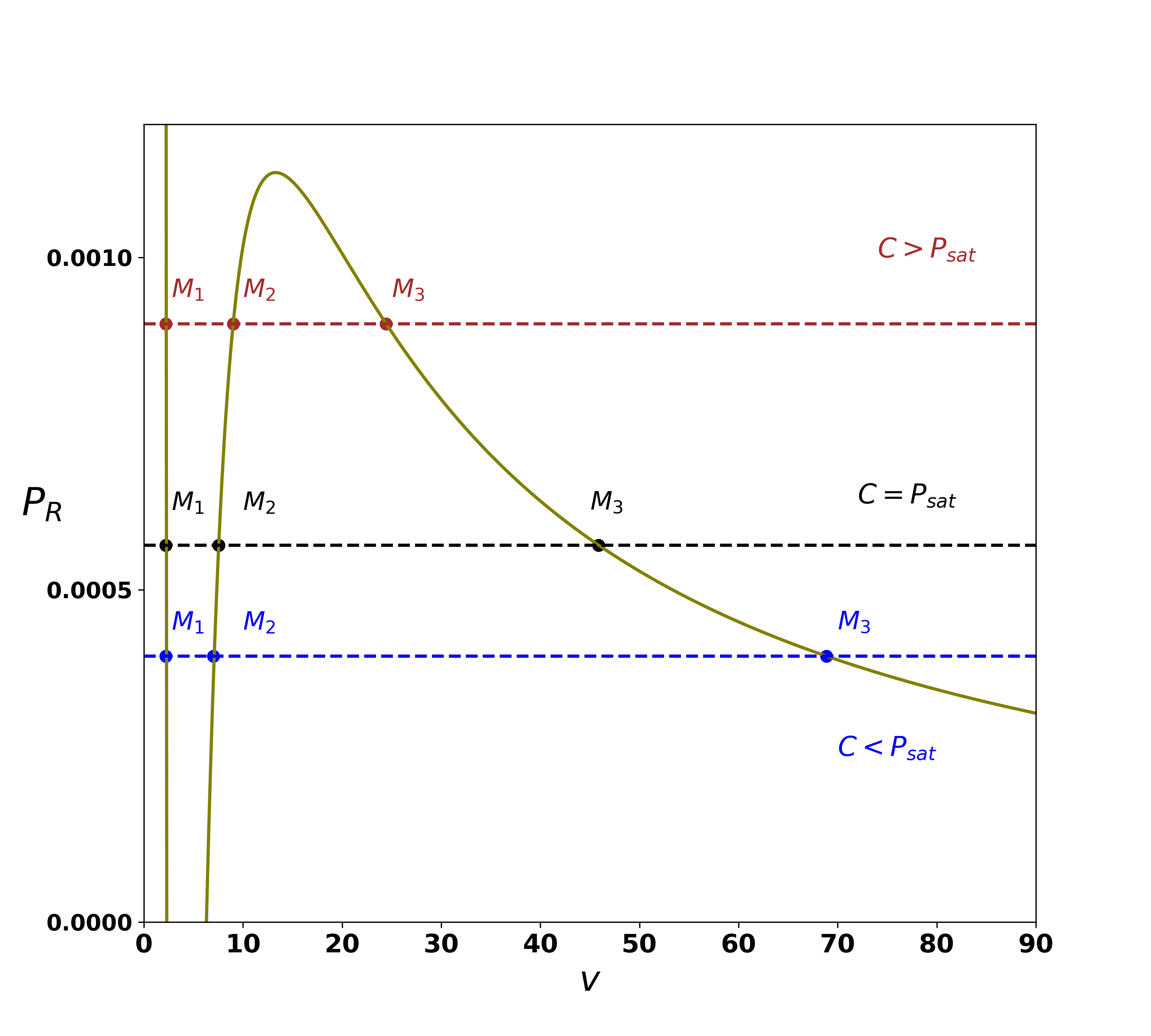}
	\end{tabbing}
 \vspace{-.8cm}
	\caption{\footnotesize\it Sub-critical Isotherms in the $P_R-v$ plane. {\bf(Left:)} for $ T_R=0.8T_{cr} $, {\bf(Right:)} for $ T_R=0.5T_{cr} $. The black dashed lines represent the saturation pressures. The fixed points for each ambient pressure are shown as $M_1$, $M_2$, and $M_3$.}
	\label{p-v_space}
\end{figure}

We now focus on investigating the thermal chaos in the Rényi charged-flat black hole, triggered by space-dependent periodic perturbations. As in the previous analysis, we consider the black hole to be in an equilibrium state at a sub-critical temperature, $T_R = T_0 \leq T_{cr}$. Drawing from the Van der Waals–Korteweg theory\cite{waals_kortweg1979,Row1979waals_korteweg}, we examine the behavior of the Cauchy stress tensor in the absence of fluid flow:
\begin{equation}\label{piola_space}
\tau=\displaystyle-P(v,T_0)-A\partial_{xx}v.
\end{equation}
In a static equilibrium state devoid of body forces, the momentum balance equation, Eq.\eqref{conserv_eq2}, implies that $\partial_x \tau = 0$. This condition establishes that $\tau = C = \text{constant}$. Here, $C$ represents the ambient pressure surrounding the fluid tube, that is, the pressure at its two end sections. Consequently, the governing dynamical equation for the black hole fluid becomes:
\begin{equation}\label{master_eq_space}
\displaystyle-P(v,T_0)-A\partial_{xx}v=C.
\end{equation}
Taking the form of a space-dependent periodic thermal perturbation of the static equilibrium such as
\begin{equation}
T=T_0+\epsilon \cos(\xi x),
\end{equation}
where $\xi$ is its spacial wave number. Substituting this term in the dynamical equation \eqref{master_eq_space}, one can achieve
\begin{equation}\label{master_eq_space_ptrb}
\displaystyle-P(v,T_0)-A\partial_{xx}v-\frac{\epsilon \cos(\xi x)}{v}=C.
\end{equation}
%{\bf Here XXXXX}
For any given sub-critical temperature $ T_0 < T_{cr} $, the dynamical system described by Eq.\eqref{master_eq_space_ptrb} presents three fixed points, corresponding to the intersections between the isotherm $ T_R=T_0 $ and the isobar $ P_R=C $ in the $P_R-v$ diagram. This scenario is illustrated in Fig.\ref{p-v_space}, where two sub-critical temperatures are depicted: $ T_R=0.8T_{cr} $ on the left and $ T_R=0.5T_{cr} $ on the

\begin{figure}[ht!]

			\begin{tabbing}
   \hspace*{-1.5cm}
			% \hspace{9.3cm}
   % \=\kill
   \includegraphics[width=9cm,height=7cm]{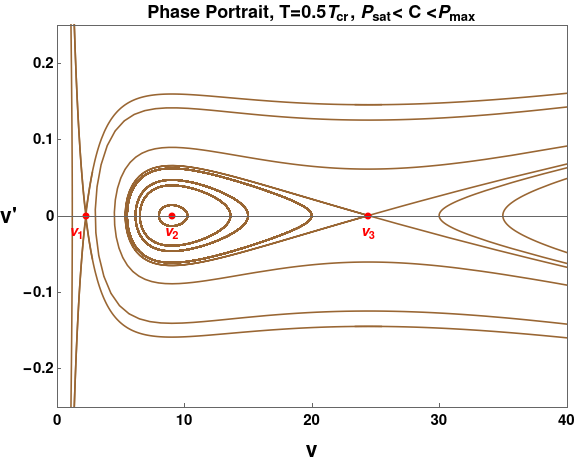}
   % \>
		\includegraphics[width=9cm,height=7cm]{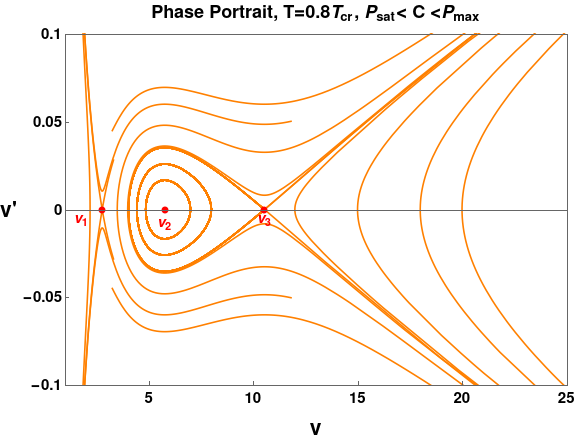}\\
     \hspace*{-1.5cm}
  \includegraphics[width=9cm,height=7cm]{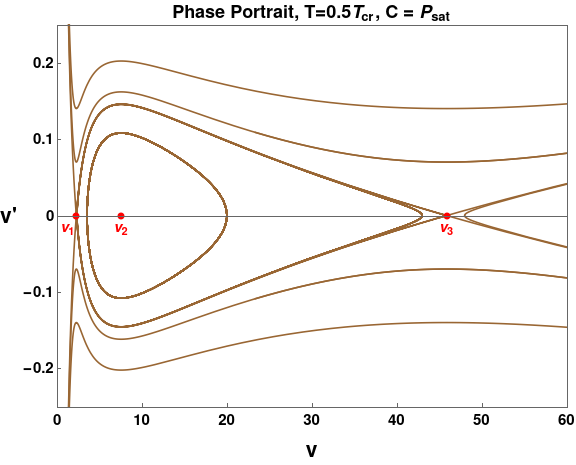}
  % \>
		\includegraphics[width=9.3cm,height=7.3cm]{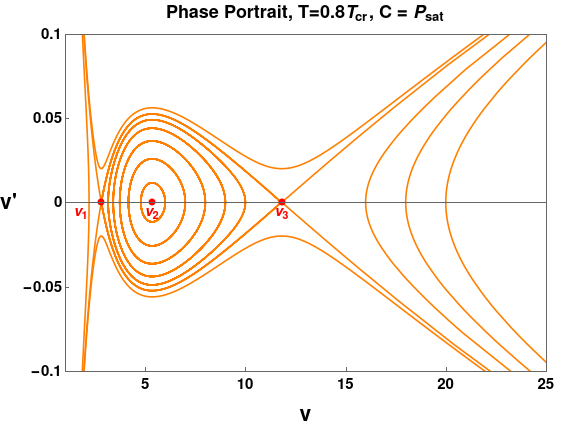}\\
     \hspace*{-1.5cm}
  \includegraphics[width=9cm,height=7cm]{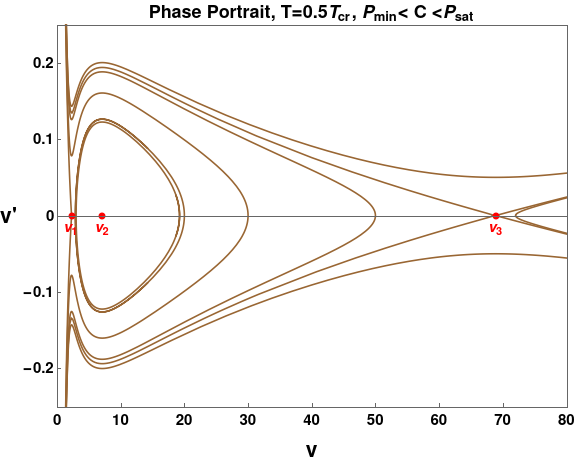}
  % \>
		\includegraphics[width=9cm,height=7cm]{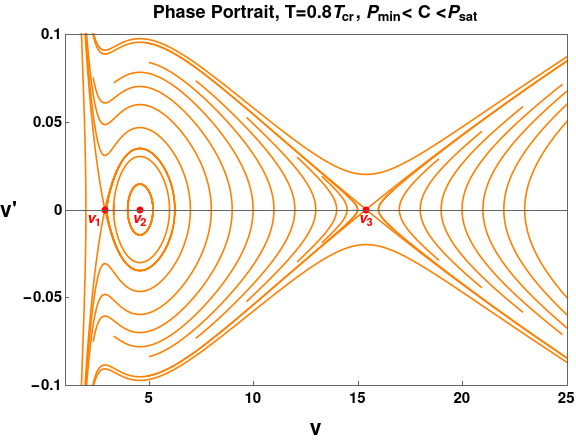}
	\end{tabbing}\vspace{-.8cm}
	\caption {\footnotesize\it $v' - v$ phase portrait of the unperturbed static black hole solutions. The (Top), (Middle) and (Bottom) panels correspond to the cases with $ P_{sat} < C < P_{max} $, $C=P_{sat}$ and $ P_{min} < C < P_{sat}$. Temperature is fixed at $T_0=0.5T_{cr}$ {\bf (Left)} and $T_0=0.8T_{cr}$ {\bf (Right)}.} 
	\label{fig:phase_portrait_space_0.8Tc}
\end{figure}

% \begin{figure}[ht!]
% \hspace{-1.5cm}
% 			\begin{tabbing}
% 			\hspace{9.3cm}\=\kill\includegraphics[width=9.5cm,height=8.5cm]{PRenyi_v_0.8Tc}\>
% 		\includegraphics[width=9.5cm,height=8.5cm]{PRenyi_v_0.5Tc}
% 	\end{tabbing}
%  \vspace{-.8cm}
% 	\caption{\footnotesize\it Sub-critical Isotherms in the $P_R-v$ plane. (a) for $ T_R=0.8T_{cr} $, (b) for $ T_R=0.5T_{cr} $. The black dashed lines represent the saturation pressures. The fixed points for each ambient pressure are shown as $M_1$, $M_2$, and $M_3$.}
% 	\label{p-v_space}\ref{}
% \end{figure}

right. The fixed points generated for ambient pressures above (brown line), below (blue line), and at the saturation pressure (black line) are denoted as $ M_1 $, $ M_2 $ and $ M_3 $, respectively. They are arranged such that $ v_1 < v_2 < v_3 $. In addition, 
the second-order dynamical equation Eq.\eqref{master_eq_space_ptrb} can be reformulated into a system of two first-order coupled differential equations as follows
\begin{align}
&\partial_x v=v',\label{dyn_sys_space1}\\
&v''=\displaystyle \frac{C-P(v,T_0)}{A}-\frac{\epsilon \cos(\xi x)}{Av}.\label{dyn_sys_space2}
\end{align}
The pressure $P(v,T_0)$ is expanded around the inflection point $v=v_0$ according to Eq.\eqref{pressure_expansion}. The general solutions that describe the {\it homoclinic} or {\it heteroclinic} orbits are obtained to be
\begin{equation}\label{key}
\theta(x)=\begin{Bmatrix}v(x)\\v'(x)\end{Bmatrix},
\end{equation}
and the dynamical equation is put under the compact form,
\begin{equation}\label{key}
\frac{d}{d t} \theta{\left(x \right)}=f(\theta)+\epsilon h(\theta,x).
\end{equation}
In which, the functions $f(\theta)$ and $h(\theta)$ are defined respectively as
\begin{equation}\label{f_g_space}
f(\theta)=\displaystyle
\begin{Bmatrix}\displaystyle v'\\\displaystyle \frac{C-P(v,T_0)}{A}
\end{Bmatrix} \quad and \quad
h(\theta,x)=
\begin{Bmatrix}\displaystyle 0\\ \displaystyle-\frac{\epsilon \cos(\xi x)}{Av}
\end{Bmatrix} .
\end{equation}
We display in Fig.\ref{fig:phase_portrait_space_0.8Tc} the unperturbed phase portraits ($\epsilon=0$) of the Rényi charged black hole fluid at three different ambient pressures and two sub-critical temperatures. The diagrams illustrate the following $T_0=0.8T_{cr}$ and $T_0=0.5T_{cr}$ cases. 
One can notice that:
\begin{itemize}

\item In all cases $v_2$ is a center point, while $v_1$ and $v_3$ are saddle points.
\item    In cases where the ambient pressure exceeds the saturation pressure ($C > P_{sat}$, top panels), a homoclinic orbit is observed. This orbit connects the saddle point at $v = v_3$ back to itself, forming a closed loop.
\item    At the saturation pressure ($C = P_{sat}$, middle panels), the system displays a heteroclinic orbit. This orbit connects the two saddle points located at $v_1$ and $v_3$.
\item    When the ambient pressure falls below the saturation pressure ($C < P_{sat}$, bottom panels), a homoclinic orbit emerges again. This time, it connects the saddle point at $v_1$ back to itself, completing the cycle.

\end{itemize}

These three distinct types of phase structures in the $v'-v$ plane observed for the Rényi charged-flat black hole mirror those seen in the charged-AdS \cite{Chabab:2018lzf}, Born–Infeld-AdS \cite{Chen_Li_Zhang_2019}, Gauss-Bonnet \cite{Mahish:2019tgv} and a variety of other black hole configurations \cite{Tang:2020zhq,Dai:2020wny}. This similarity suggests a common characteristic among static AdS black holes and supports the proposed link between these AdS black holes described within Gibbs-Boltzmann statistics and flat black holes analyzed through the Rényi formalism.
%\newpage
\subsection{Mel'nikov function for space-dependent perturbations}
Continuing our exploration of chaos, the Mel'nikov function in this context is space-dependent and is defined as follows
\begin{equation}\label{Melnk_space_def}
\mathbf{\mathcal{M}}(x_0)=\int_{-\infty}^{\infty}f^T(\theta_0(x))\;\mathcal{J}\;h(\theta_0(x),x)\bigg\rvert_{x=\tilde{x}-x_0}d\tilde{x},
\end{equation}
where $\theta_0(x)=\begin{Bmatrix}v_0(x)\\v'_0(x)\end{Bmatrix}$ is the unperturbed solution associated with the vanishing $ \epsilon $ situation and is governed by the dynamical system described in Eq.\eqref{dyn_sys_space1} and Eq.\eqref{dyn_sys_space2}. The functions $f$ and $h$ are specified in Eq.\eqref{f_g_space}. Consequently, after simplifications, the Mel'nikov function is evaluated to be
\begin{equation}\label{Melnk_space_exp}
\mathbf{\mathcal{M}}(x_0)=-\int_{-\infty}^{\infty}\frac{u_0(\tilde{x}-x_0)\cos[\xi(\tilde{x}-x_0)]}{Av_0(\tilde{x}-x_0)}d\tilde{x}.
\end{equation}
As before it is straightforward to put the Mel'nikov function in the usual form
\begin{equation}\label{Melnk_space_sol}
\mathbf{\mathcal{M}}(x_0)=-\mathcal{L'}\cos(\xi x_0)+\mathcal{K'}\sin(\xi x_0).
\end{equation}
With,
\begin{equation}\label{key}
\mathcal{L'}=\int_{-\infty}^{\infty}\frac{u_0(\tilde{x}-x_0)cos(\xi \tilde{x})}{v_0(\tilde{x}-x_0)}d\tilde{x}\quad and \quad
\mathcal{K'}=\int_{-\infty}^{\infty}\frac{u_0(\tilde{x}-x_0)cos(\xi \tilde{x})}{v_0(\tilde{x}-x_0)}d\tilde{x}.
\end{equation}
The form of the Mel'nikov function, Eq.\eqref{Melnk_space_sol} signifies that it possesses simple zeros for arbitrary values of $\mathcal{L'}$, $\mathcal{K'}$ and $\xi$. 
\begin{equation}\label{Melnk_space_sol-2}
\mathbf{\mathcal{M}}(x_0)=0\implies  x_0=\frac{1}{\xi}\arctan\left(\frac{\mathcal{L'}}{\mathcal{K'}}\right),
\end{equation}

\begin{figure}[ht!]

			\begin{tabbing}
   \hspace*{-1.6cm}
			% \hspace{9.3cm}
   % \=\kill
   \includegraphics[width=9.3cm,height=6.3cm]{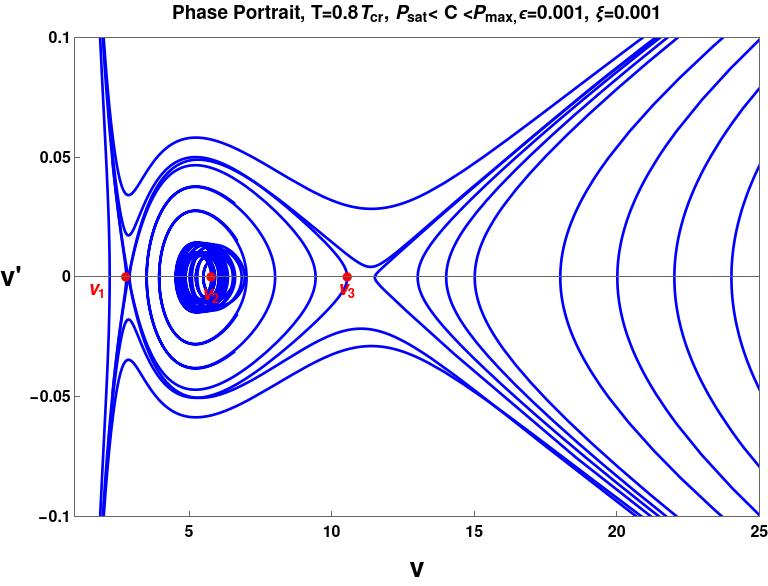}
	% \>
		\includegraphics[width=9.3cm,height=6.3cm]{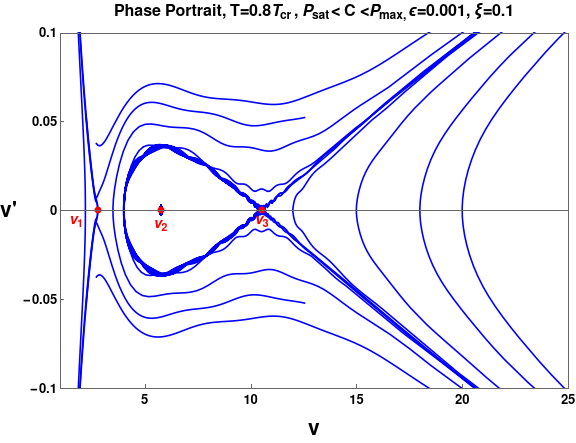}
	\\
	\hspace*{-1.6cm}
		\includegraphics[width=9.3cm,height=6.3cm]{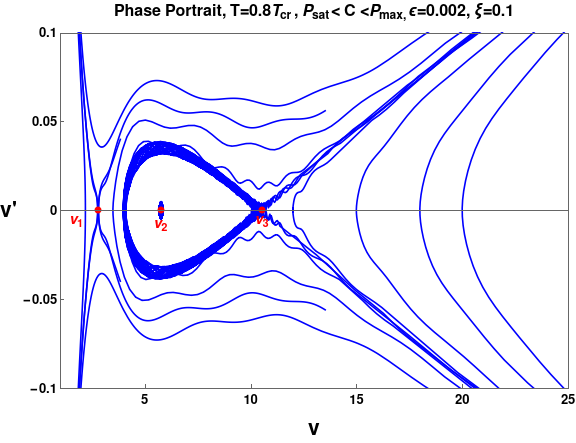}
	% \>
		\includegraphics[width=9.3cm,height=6.3cm]{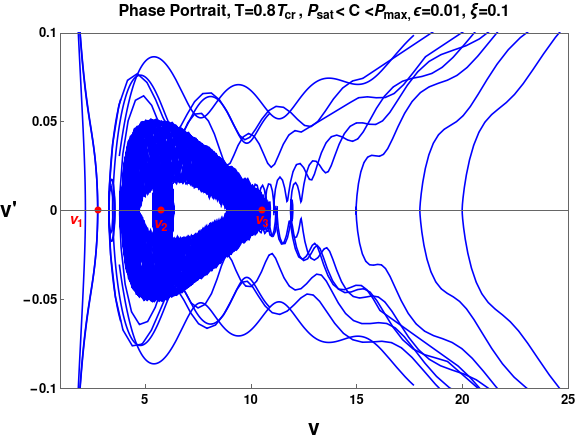}
	\\
 \end{tabbing}
   \vspace{-0.8cm}
	\caption{\footnotesize\it  $v' - v$ phase portrait of the spatially perturbed static black hole solution. The panels correspond to the case with $ P_{sat} < C < P_{max} $ and temperature is fixed at $T_0=0.8T_{cr}$.}
	\label{fig:phase_portrait_pert_space}
\end{figure}

This also means that chaos is ubiquitous for the charged-flat black hole under space-dependent periodic thermal perturbation. In Fig.\ref{fig:phase_portrait_pert_space}, we plotted the solutions of the perturbed equation Eq.\eqref{master_eq_space_ptrb} in the phase plane.
It is clear that spatial chaos appears irrespective of the perturbation parameters. An identical conclusion on the ubiquity of space-dependent chaos was reached also in the case of charged-AdS black holes.

%The analysis reveals that spatial chaos emerges regardless of the perturbation parameters involved.
%\newpage
\section{Conclusion}
In this work, we have conducted an analytical investigation into the chaotic dynamics of charged-flat black holes within an extended phase space constructed via Rényi statistics. Employing the well-established Poincaré-Mel'nikov theory, we computed the Mel'nikov function and identified its zeros, demonstrating the presence of temporal perturbations in the spinodal region. Additionally, our analysis revealed instances of spatial chaos in the equilibrium configurations of small and large black holes.

%We've explored the influence of thermal chaos in the Rényi extended phase space of a charged flat black hole, applying the Mel'nikov method to assess how the charge parameter impacts this phenomenon. 

In addition, our findings reveal that in the presence of temporal perturbations within the spinodal region, chaos manifests only if the amplitude of perturbation, $\delta$, surpasses a critical level, $\delta_c$. Moreover, this study also revealed the important stabilizing influence of the electric charge against time-periodic thermal fluctuations as well as an unexpected resonant behavior shown by the black hole fluid. This threshold behavior aligns with what is observed in similar anti-de-Sitter systems like Reissner-Nordstrom, Born–Infeld, dilatonic, and Gauss-Bonnet black holes.  For spatially periodic thermal perturbations, chaos is consistently present, contrasting with the conditional chaos triggered by temporal perturbations. This consistent emergence of chaos in spatial perturbations, irrespective of amplitude, is a characteristic observed among static AdS black holes, and can also be observed in asymptotically flat black hole configurations within the Rényi statistics. This analogy hints at a potential deeper connection between the cosmological constant $\Lambda$ and the nonextensivity Rényi parameter $\lambda$, setting the stage for further investigation.

%\appendix
%\numberwithin{equation}{section}
%\newpage

\bibliographystyle{unsrt}
\bibliography{aRChaos} 
\end{document}